\documentclass[12pt]{iopart}

\expandafter\let\csname equation*\endcsname\relax
\expandafter\let\csname endequation*\endcsname\relax

\usepackage{amsmath,amsthm,amsfonts,amssymb,bm,graphicx,color,times, braket}
\usepackage[colorlinks={true}, citecolor={blue}, filecolor={blue}, linkcolor={blue}, urlcolor={blue}]{hyperref}
\usepackage[caption=false]{subfig}
\usepackage[T1]{fontenc}
\usepackage[latin9]{inputenc}
\usepackage{lmodern} 
\usepackage{color}
\usepackage{bbold}
\usepackage{dsfont}
\usepackage{graphicx}
\usepackage{enumerate}
\usepackage{float}
\usepackage{mathtools}
\usepackage{leftindex}
\usepackage{cite}

\usepackage{empheq}
\usepackage{tikz}
\usepackage[normalem]{ulem}
\DeclareMathAlphabet\mbc{OMS}{cmsy}{b}{n}
\usepackage[latin9]{inputenc}

\usepackage{empheq}
\usepackage{tikz}
\usepackage[normalem]{ulem}
\DeclareMathAlphabet\mbc{OMS}{cmsy}{b}{n}
\usepackage{balance}

\usepackage{soul}

\global\long\def\ket#1{\left|#1\right\rangle }
\global\long\def\bra#1{\left\langle #1\right|}

\global\long\def\sbkt#1{\left[#1\right]}
\global\long\def\cbkt#1{\left\{#1\right\}}

\global\long\def\re{\mathrm{Re}}
\global\long\def\im{\mathrm{Im}}

\global\long\def\mr#1{\mathrm{#1}}
\global\long\def\mb#1{{\mathbf #1}}

\global\long\def\non{\nonumber}

\allowdisplaybreaks

\begin{document}

\title[]{Steady-State Entanglement Generation via Casimir-Polder Interactions}
\author{Mohsen Izadyari}

\address{College of Optical Sciences and Department of Physics, University of Arizona,  Tucson, AZ, 85721, USA}
\address{Department of Physics, Ko\c{c} University, 34450 Sar{\i}yer, Istanbul, T\"{u}rkiye}
\address{Faculty of Engineering and Natural Sciences, Kadir Has University, 34083 Fatih, Istanbul, T\"{u}rkiye}
\ead{mizadyari18@ku.edu.tr}
\author{Onur Pusuluk}
\address{Faculty of Engineering and Natural Sciences, Kadir Has University, 34083 Fatih, Istanbul, T\"{u}rkiye}
\ead{onur.pusuluk@gmail.com}

\author{Kanu Sinha}
\ead{kanu@arizona.edu}
\address{College of Optical Sciences and Department of Physics, University of Arizona,  Tucson, AZ, 85721, USA}
\author{\"{O}zg\"{u}r E. M\"{u}stecapl{\i}o\u{g}lu}
\ead{omustecap@ku.edu.tr}
\address{Department of Physics, Ko\c{c} University, 34450 Sar{\i}yer, Istanbul, T\"{u}rkiye} 
\address{T\"{U}B\.{I}TAK Research Institute for Fundamental Sciences, 41470 Gebze, T\"{u}rkiye}
\address{Faculty of Engineering and Natural Sciences, Sabanci University, 34956 Tuzla, Istanbul, T\"{u}rkiye}
\vspace{10pt}
\begin{indented}
\item[]\today
\end{indented}

\begin{abstract}
We investigate the generation of steady-state entanglement between two atoms resulting from the fluctuation-mediated Casimir-Polder (CP) interactions near a surface. Starting with an initially separable state of the atoms, we analyze the atom-atom entanglement dynamics for atoms placed at distances in the range of $\sim25$~nm away from a planar medium, examining the effect of medium properties and geometrical configuration of the atomic dipoles. We show that perfectly conducting and superconducting surfaces yield an optimal steady-state concurrence value of approximately 0.5. Furthermore, although the generated entanglement decreases with medium losses for a metal surface, we identify an optimal distance from the metal surface that assists in entanglement generation by the surface. While fluctuation-mediated interactions are typically considered detrimental to the coherence of quantum systems at nanoscales, our results demonstrate a mechanism for leveraging such interactions for entanglement generation.
\end{abstract}

%
%
%
%
%

\section{Introduction}
Nanoscale quantum systems enable efficient and tunable light-matter interactions by confining EM fields in small mode volumes~\cite{DamicoNQO, ChangRMP2018,Atom-field003,Atom-field004,Atom-field005,atom-field007}. Such systems hold enormous promise in the development of photonic devices that exhibit nonlinearities at the single-photon level~\cite{QNonLinear01,Nonlinear002,Nonlinear003,Skljarow22}, building efficient light-matter interfaces \cite{Goban14,coupling011,interface001,interface002,coupling22,atom-field006} that are essential for various quantum information processing tasks \cite{QInf001,QInf002}, as well as, exploring fundamental phenomena such as ultrafast dynamics at the nanoscale \cite{Ultra001,Ultra002}.

However, as quantum systems move towards nanoscales, while the coupling between atomic systems and the relevant modes of the EM field increases, so does the detrimental interaction of atomic systems with the quantum fluctuations of the EM field \cite{dipole-dipole001}. Such an interaction between a neutral macroscopic body and polarizable particles via the quantum fluctuations of the EM field leads to fluctuation-mediated forces, known as Casimir-Polder (CP) forces~\cite{CasimirPolder01, BookDispersion01}. These forces, as well as, the concomitant dissipation and decoherence arising from quantum fluctuations near surfaces, critically limit trapping schemes and the coherence of quantum emitters near surfaces \cite{CPTrap01,CPTrap02, Trap-Kanu02}. 
Realizing quantum technologies requires the production and longevity of quantum coherence and entanglement as critical resources for performing quantum-enhanced tasks that rely on superposition and correlations of quantum systems~\cite{QTech001,QTech002}. 
To this end, sophisticated quantum control methods, and environmental noise engineering or noise-assisted entanglement generation schemes have been proposed~\cite{Control001,control002,control003,control004,BathEnt01,BathEnt02,BathEnt03}. 
Given that fluctuation-mediated interactions are deleterious for quantum coherence and correlations of atomic systems near surfaces \cite{ScheelMain,Cas_Ent01,Noise001,Noise002,Noise003,Noise004,Noise005,Noise006,Dipole-Dipole2024,Lambrecht2015,Donaire_2015,BathEnt04}, engineering such fluctuation phenomena to achieve better coherence and control of nanoscale quantum systems thus becomes a vital goal in developing nanoscale quantum devices \cite{QInf002}. 
The generation of entanglement between resonantly interacting atoms mediated by microsphere dielectric has been studied in Ref.~\cite{ScheelMain}, demonstrating that the entanglement arises from a medium-assisted interaction but rapidly decays from its maximum value to zero. However, optimizing the medium's optical properties and the system's geometry can potentially lead to more stable entanglement.
\par In this work, we propose a simple and intuitive scheme for controlling fluctuation phenomena to produce steady-state entanglement between two quantum emitters near a planar half-space medium. Our approach utilizes the surface-assisted dissipative interaction between two atoms in the presence of a planar medium to produce robust steady-state entanglement for critical placement of the atoms relative to the surface. The results offer a new perspective on engineering surface-mediated systems by leveraging the medium's material properties, geometry, and the spatial coordinates of the atoms relative to the surface to generate and preserve entanglement.
\begin{figure}[b]
    \centering
    \includegraphics[width=0.4\linewidth]{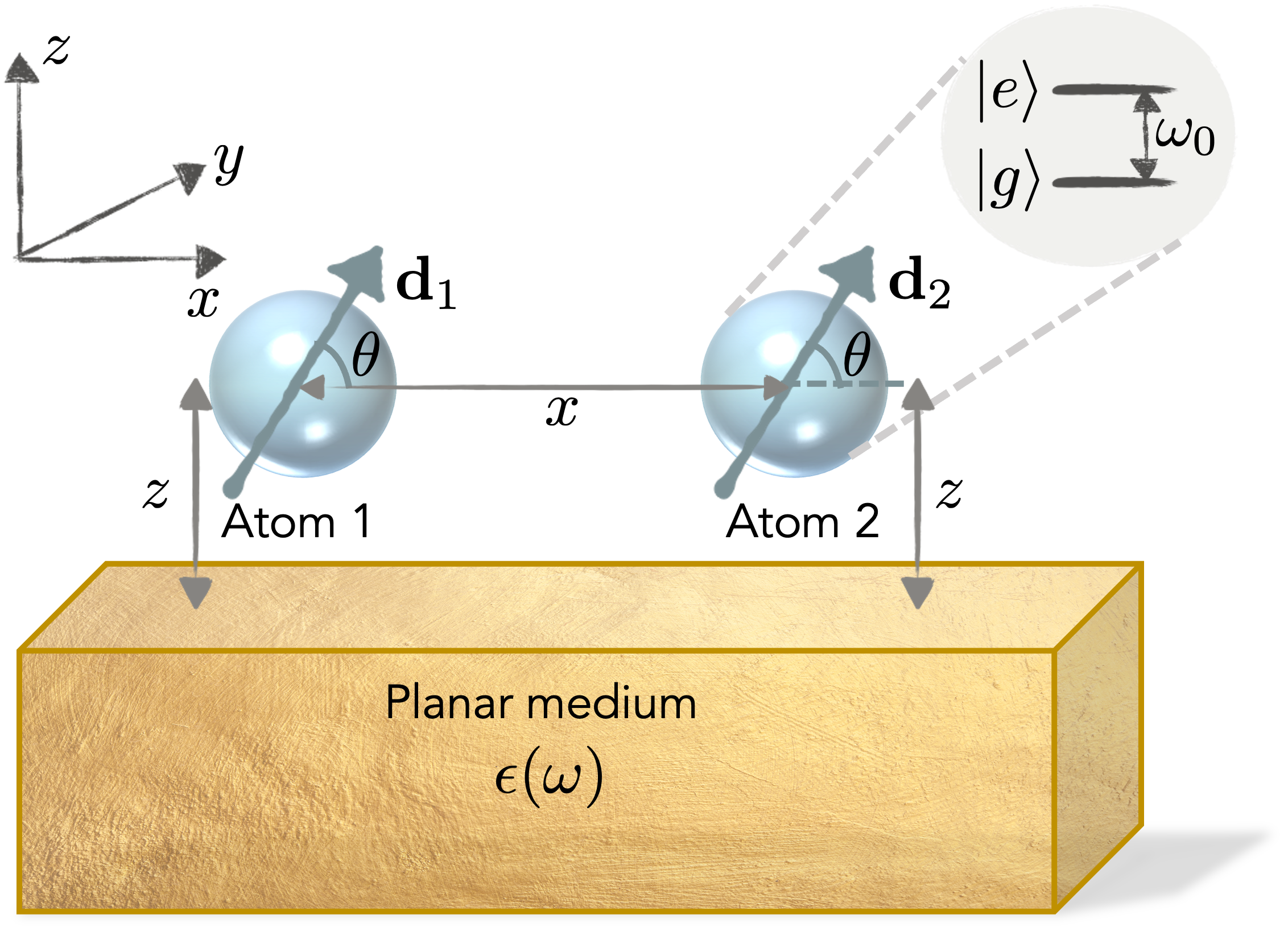}
    \caption{Schematic of the model featuring two two-level atoms near a planar medium of permittivity $ \epsilon (\omega)$. The atomic dipoles $ \mb{d}_1$ and $ \mb{d}_2$ are oriented along $\theta$ with respect to the $x$-axis.  }
    \label{fig:Model}
\end{figure}
\section{\label{sec:system}Model}

We present a system of two emitters positioned at a distance $x$ from each other and at a distance $z$ from a planar half-space medium characterized by electric permittivity $\epsilon(\omega)$ (Fig.~\ref{fig:Model}). Each emitter, denoted by $i = 1, 2$ for the first and second emitter respectively, is modeled as a two-level system with ground state $\ket{g}_i$ and excited state $\ket{e}_i$, with a resonant transition frequency $ \omega_0$. In the presence of the planar surface, the electric field originating from vacuum fluctuations at position $\mathbf{r}$ is defined as \cite{BookDispersion01}:
\begin{align}\label{Eq:Electric-field}
    \hat{E}(\textbf{r}) = \sum_{\lambda=e,m} \int d^3\mb{r}' \int d \omega ~[\Bar{G}_{\lambda}(\mb{r}',\mb{r}',\omega).\hat{f}_{\lambda}(\textbf{r}'.\omega) + \text{H.c.}]
\end{align}
Here, $\hat{f}_{\lambda} (\mb{r}, \omega) $ and $\hat{f}_{\lambda}^{\dagger} (\mb{r}, \omega)$ are bosonic annihilation and creation operators, representing noise polarization ($ \lambda = e $) and magnetization ($ \lambda = m $) excitations at frequency $\omega$. The coefficients $\bar {G}_\lambda(\mb{r}, \mb{r}', \omega) $ are related to the  Green's tensor $\Bar{G}(\mb{r}, \mb{r}', \omega)$, which satisfies the inhomogeneous Helmholtz equation in the presence of the medium (See \ref{App.A} for details). The propagator $\Bar{G}(\mb{r}, \mb{r}', \omega)= \bar{G}_{\text{free}}(\mathbf{r}, \mathbf{r}', \omega) + \bar{G}_{\text{sc}}(\mathbf{r}, \mathbf{r}', \omega)$ consists of the free space and scattering  components.
The total Hamiltonian, denoted as $\hat{H}_T = \hat{H}_S + \hat{H}_F + \hat{H}_{\text{int}}$, comprises of the atomic Hamiltonian $\hat{H}_S = \sum_{i=1}^2 \hbar  \omega_0 \hat{\sigma}_i^{+} \hat{\sigma}_i^{-}$, the surface-assisted electromagnetic field in vacuum state $\hat{H}_F$, and the interaction Hamiltonian. $\hat \sigma_{i}^{+} = (\hat \sigma_{i}^{-})^{\dagger} = \ket{e}_{i}\bra{g}_{i}$ indicate the ladder operators for the atomic transition. The interaction Hamiltonian is described by $\hat{H}_{\text{int}} = -\sum_{i=1}^2 \hat{d}_i \cdot \hat{E}(\textbf{r}_i)$, where $\hat{d}_i = \textbf{d}_i \hat \sigma_i^{+} + \textbf{d}_i^* \hat \sigma_i^{-}$ represents the electric-dipole operator associated with the $i$-th emitter located at position $\textbf{r}_i$, with $\textbf{d}_i$ denoting the electric dipole moment.  

The time evolution of the density matrix $\rho_S$ for the emitters is governed by the Born-Markov master equation~\cite{BookMaster03}:
\begin{equation} \label{eq: MasterEq}
    \frac{d \hat{\rho}_{S}}{dt} = -\frac{i}{\hbar}[\hat{\mathcal{H}}_{\text{eff}},\hat{\rho}_{S}] + \mathcal{L}_{\text{eff}}[\hat{\rho}_{S}]
\end{equation}
Here $\hat{\mathcal{H}}_{\text{eff}}$ is the atomic effective Hamiltonian in interaction picture defined by \cite{Collective-Kanu01, Olivera22, sone2024}

\begin{align}\label{Eq:EffHam}
    \hat{\mathcal{H}}_{\text{eff}} =& \hbar \sbkt{\sum_{i=1,2} \left(\Omega_{i}^{(+)}(\mathbf{r}_i) \Hat{\sigma}_i^+\Hat{\sigma}_i^- + \Omega_{i}^{(-)}(\mathbf{r}_i) \Hat{\sigma}_i^-\Hat{\sigma}_i^+\right) + \sum_{i\neq j} \Omega_{ij} (\mathbf{r}_i,\mathbf{r}_j) \Hat{\sigma}_i^- \Hat{\sigma}_j^+},
\end{align}

and the effective surface-modified Liouvillian in Eq.~(\ref{eq: MasterEq}) is given by

\begin{align}\label{Eq: Liou}
    &\mathcal{L}_{\text{eff}}[\hat{\rho}_{S}] =\sum_{ij}\frac{\Gamma_{ij}(\mathbf{r}_i,\mathbf{r}_j)}{2} \left(2\Hat{\sigma}_i^- \hat{\rho}_{S} \hat{\sigma}_j^+ - \hat{\sigma}_i^+ \hat{\sigma}_j^-\hat{\rho}_{S} - \hat{\rho}_{S}\hat{\sigma}_i^+ \hat{\sigma}_j^- \right), 
\end{align}
where $\Gamma_{ij} (\mathbf{r}_i,\mathbf{r}_j)= \Gamma_{ij}^{\text{free}}(\mathbf{r}_i,\mathbf{r}_j) + \Gamma_{ij}^{\text{sc}}(\mathbf{r}_i,\mathbf{r}_j)$ corresponds to the dissipative coupling coefficient between two atoms ($i\neq j$) accounts for the near-surface cooperative decay, while $\Gamma_{ii}$ represents the spontaneous emission rate for the excited state of the $i$th atom~\cite{Collective-Kanu01} (more discussions in \ref{App.C}).

The diagonal elements of the effective Hamiltonian $\hat{\mathcal{H}}_{\text{eff}}$ are the CP shift contributions defined as:

\begin{align} \label{Eq:CPShift}
    \Omega_{i}^{(-)} (\mathbf{r}_i) &= \frac{\mu_0 \omega_0}{\pi} \int_0^{\infty} \frac{d\zeta~\zeta^2}{\zeta^2 + \omega^2}\mathbf{d}_i^*\cdot [ \Bar{G}_{\mr{\text{sc}}}(\mathbf{r}_i,\mathbf{r}_i,i\zeta)] \cdot \mathbf{d}_i \nonumber \\
    \Omega_{i}^{(+)}(\mathbf{r}_i) &= -\Omega_{i}^{(-)}(\mathbf{r}_i) + \Omega_i^{\mr{\text{res}}}(\mathbf{r}_i),
\end{align}
where, the free component of Green's tensor is neglected, assuming it has already been taken into account as a contribution to Lamb shifts in the bare levels. Additionally, the resonant contribution $\Omega_i^{\mr{\text{res}}}(\mathbf{r}_i) = -\mu_0 \omega_0^2 \re[\mathbf{d}_i^*\cdot \Bar{G}_{\mr{\text{sc}}}(\mathbf{r}_i,\mathbf{r}_i,\omega_0) \cdot \mathbf{d}_i]$, is dependent exclusively on the environment's response at the resonant frequency where $\mu_0$ is the Permeability of free space.

The off-diagonal elements $(i \neq j)$ of the effective Hamiltonian, along with the dissipative coupling coefficients, are determined by the resonant contribution of the real and imaginary parts of Green's function. This is expressed as $\Omega_{ij} (\mathbf{r}_i,\mathbf{r}j) = \Omega_{ij}^{\mr{\text{free}}} (\mathbf{r}_i,\mathbf{r}j) + \Omega_{ij}^{\mr{\text{sc}}} (\mathbf{r}_i,\mathbf{r}_j)$ as follows:

\begin{align}\label{Eq: O12}
    \Omega_{ij}^{\mr{\text{free},\text{sc}}} (\mathbf{r}_i,\mathbf{r}_j) =&-\mu_0 \omega_0^2 \re[\mathbf{d}_i^*\cdot \Bar{G}_{\mr{\text{free},\text{sc}}}(\mathbf{r}_i,\mathbf{r}_j,\omega_0) \cdot \mathbf{d}_j],\nonumber \\
    \Gamma_{ij}^{\mr{\text{free},\text{sc}}} (\mathbf{r}_i,\mathbf{r}_j) =& 2\mu_0 \omega_0^2~\im[\mathbf{d}_i^*\cdot \Bar{G}_{\mr{\text{free},\text{sc}}}(\mathbf{r}_i,\mathbf{r}_j,\omega_0) \cdot \mathbf{d}_j].
\end{align}
\section{Results} 
We analyze the dynamics of the emitters, governed by the master equation~\eqref{eq: MasterEq}.
As illustrated in Fig.~\ref{fig:Model}, we specify $x_1 - x_2 \equiv x$ and note that the two atoms are positioned at the same distance $z$ from the surface. Consequently, in accordance with Eqs.~\eqref{Eq:CPShift} and \eqref{Eq: O12}, we can simplify the system by setting $\Omega_1^{(\mp)}(z) = \Omega_2^{(\mp)}(z) \equiv \Omega^{(\mp)}$, $\Omega_{12}^{(\mp)}(x,z) = \Omega_{21}^{(\mp)}(x,z) \equiv \Omega_{12}^{(\mp)}$, $\Gamma_{11}(z) = \Gamma_{22}(z) \equiv \Gamma$, and $\Gamma_{12}(x,z) = \Gamma_{21}(x,z) \equiv \Gamma_{12}$.

We introduce the atomic density matrix as $ \hat \rho_S = \rho_{11} \ket{ee}\bra{ee} + \rho_{22} \ket{eg}\bra{eg} + \rho_{33} \ket{ge}\bra{ge} + \rho_{23} \ket{eg}\bra{ge} + \rho_{32} \ket{ge}\bra{eg} + \rho_{44} \ket{gg}\bra{gg}$. Considering the initial state of the two atoms as $\rho_S(t_0) = \ket{eg}\bra{eg}$, the time-dependent elements of the density matrix can be obtained by solving the master equation (See \ref{App.B})
\begin{subequations}
\renewcommand{\theequation}{\theparentequation.\arabic{equation}}
    \begin{align}\label{Eq:EqOfMotion02}
    \rho_{11}(t) &= 0, \\
     \rho_{44}(t) &= 1 - \frac{1}{2}(e^{-(\Gamma - \Gamma_{12})t} + e^{-(\Gamma + \Gamma_{12})t}), \label{eq:2.2}\\
     \Phi_{+}(t) &= \frac{1}{2}(e^{-(\Gamma - \Gamma_{12})t} + e^{-(\Gamma + \Gamma_{12})t}), \label{eq:2.3}\\
     \Phi_{-}(t) &= \cos{(2 \Omega_{12} t)}e^{-\Gamma t},    \\
       \Psi_{+}(t) &= - \frac{1}{2}(e^{-(\Gamma - \Gamma_{12})t} - e^{-(\Gamma + \Gamma_{12})t}),  \label{eq:2.5} \\
       \Psi_{-}(t) &= i\sin{(2 \Omega_{12}t)}e^{-\Gamma t} \label{eq:2.6}
    \end{align}
\end{subequations}
Here, we define $\Phi_{\pm} \equiv \rho_{22} \pm \rho_{33}$ and $\Psi_{\pm} \equiv \rho_{23} \pm \rho_{32}$. 
According to Eqs. \eqref{Eq:EqOfMotion02}--\eqref{eq:2.6}, all elements of the density matrix decay to zero as $t \rightarrow \infty$, except for $\rho_{44}$ (see Eq.~\eqref{eq:2.2}), which signifies both atoms being in their ground state, as expected. However, when $\Gamma = \Gamma_{12}$ (Eqs.~\eqref{eq:2.3}) and (\ref{eq:2.5})), the system evolves to the steady state:

\begin{align}\label{Eq:steadystate02}
    \rho_{\text{steady}} = \frac{1}{2} \left(\ket{\psi_{sub}}\bra{\psi_{sub}} + \ket{gg}\bra{gg}\right),
\end{align}
where $\ket{\psi_{\text{{sub}}}} \equiv \left(\ket{eg} - \ket{ge}\right)/\sqrt{2}$ corresponds to the subradiant state of the emitters that is decoupled from the EM field~\cite{Dicke}.  Thus we observe that quantum correlations emerge between the two emitters, transforming the initially uncorrelated state into an entangled steady state, with a concurrence of $ C(\rho_\mr{steady}) = 0.5$~, which quantifies the degree of entanglement $(0<C<1)$ \cite{concurrence01,concurrence02,ConcurrenceRef}.

The condition $\Gamma = \Gamma_{12}$ that yields the entangled steady-state signifies a balance between the emitters' spontaneous emission rate ($\Gamma = \Gamma_{11} = \Gamma_{22}$) and the surface-mediated dipole-dipole dissipative interaction ($\Gamma_{12}$), as was also analyzed in \cite{ScheelMain}. We introduce the relative decay $D(\mathbf{r}_1,\mathbf{r}_2) = \Gamma - \Gamma_{12}$, which is intricately dependent on the spatial arrangement of two emitters, their dipole orientation, and the properties of the surface. Initiating the system with one emitter in the excited state allows for the potential delocalization of a photon between the two emitters. 

The steady-state entanglement condition thus corresponds to $D(\mathbf{r}_1,\mathbf{r}_2)$ approaching 0. Accordingly, the concurrence of the atomic system decreases as the relative decay increases (see Fig. C3 in \ref{App.C}).
We analyze two distinct dipole configurations, namely $zz$ and $xx$ (where atomic dipoles are perpendicular ($\theta=\pi/2$) and parallel ($\theta=0$) to the surface, respectively (see Fig.~\ref{fig:Model})), and investigate the behavior of $D(\mathbf{r}_1,\mathbf{r}_2)$ as a function of the emitters' position. 


The relative decay can be expressed as a sum of contributions from the free and scattering components: $D(\tilde{x},\tilde{z}) = D^{\text{free}}(\tilde{x}) + D^{\text{sc}}(\tilde{x},\tilde{z})$ where the scaled parameters are introduced as $\tilde{x} \equiv k_0 x$ and $\tilde{z} \equiv k_0 z$, with $k_0 = \omega_0/c$. The free part characterizes the system in the absence of the surface, while the scattering part accounts particularly for the surface effects. 
Utilizing Eq.~\eqref{Eq: O12}, the free part is calculated as $D^{\text{free}}(\tilde{x}) = \Gamma_0 - \Gamma_{12}^{\text{free}}(\tilde{x})$.
We note that the free part of the relative decay for both configurations as a function of $\tilde{x}$ starts from zero and tends to $\Gamma_0$ for long distances, where the dipole-dipole interaction $\Gamma_{12}^\mr{free}$ vanishes as the dipoles are infinitely separated from each other (see \ref{App.C}). 
\begin{figure}[t]
    \centering
    \includegraphics[width=0.5\linewidth]{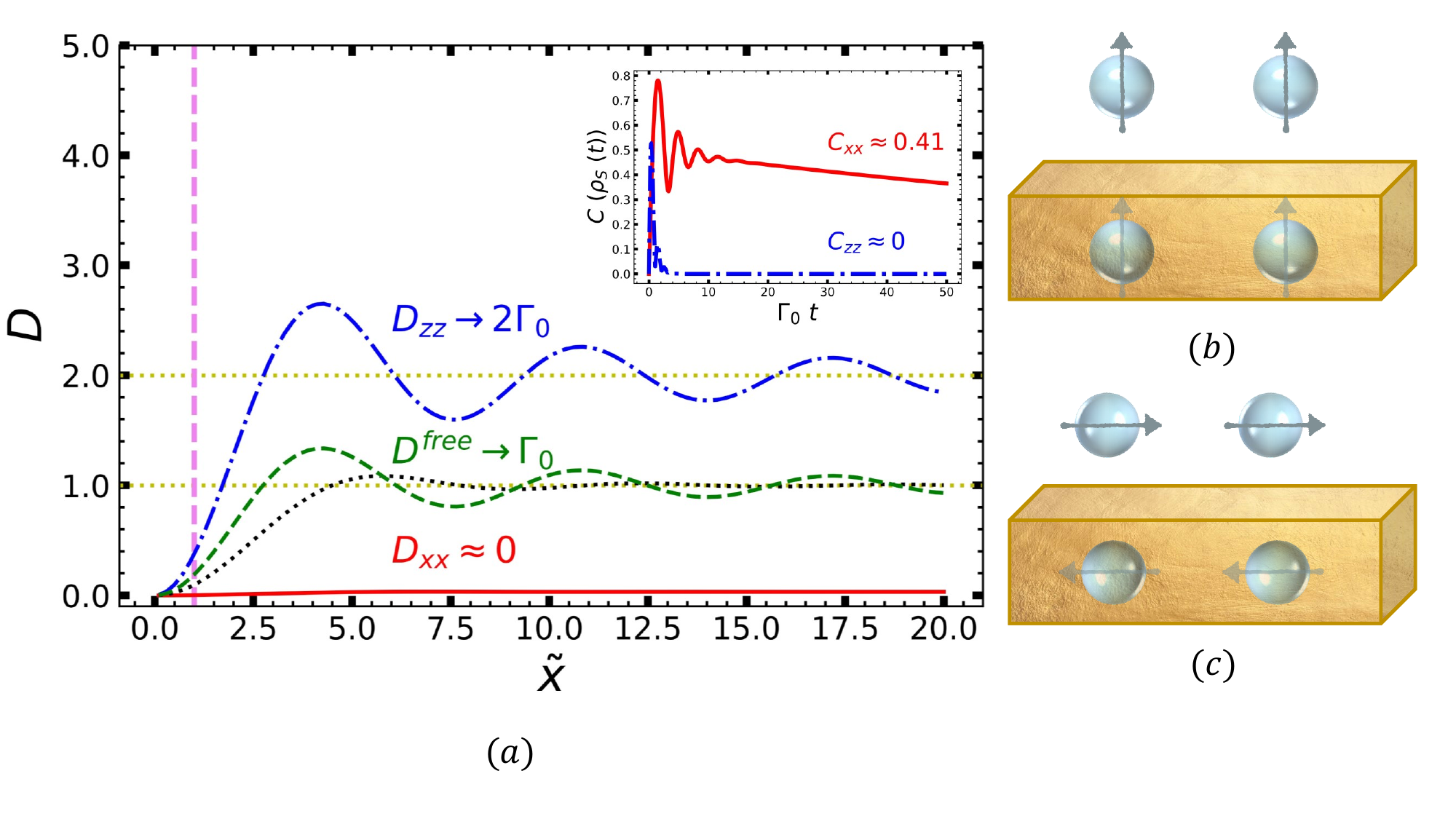}
    \caption{(a) The relative decay in free space (green dashed and black dotted curves) and the total relative decay near the perfect conductor at a distance of $\tilde{z} = 0.2$, as functions of $\tilde{x}$ for the $zz$ (dash-dotted blue curve) and $xx$ (solid red curve) configurations. The inset shows the density matrix's time-dependent concurrence for the system with $zz$ (dash-dotted blue curve) and $xx$ (solid red curve) configurations near the perfect conductor surface. (b) and (c), The schematic depicts the dipole-image model for the perpendicular dipoles in $zz$ configuration and the parallel dipoles in $xx$ configuration, along with their corresponding images in the medium.}
    \label{fig:DD_ff}
\end{figure}
The scattering part of the relative decay, besides $\tilde{x}$ and dipole orientations, relies on $\tilde{z}$ and the medium properties, defined as $D^{\text{sc}}(\tilde{x},\tilde{z}) = \Gamma^{sc}(\tilde{x},\tilde{z}) - \Gamma_{12}^{sc}(\tilde{x},\tilde{z})$.
First, we consider a perfect conductor surface with permittivity $\epsilon(\omega) \rightarrow \infty$, which is calculated for each dipole configuration based on Eq.~(\ref{Eq: O12}). According to calculations, as $\tilde{x}$ increases, the free and scattering components of the relative decay amplify each other in the $zz$ configuration, leading to $D_{zz} \rightarrow 2\Gamma_0$, while in the $xx$ configuration, their interaction is destructive, resulting in $D_{xx} \rightarrow 0$. The relative decay as a function of $\tilde{x}$ is shown in Fig.~\ref{fig:DD_ff} (a) for each dipole configuration: in free space, denoted by $D_{zz}^\mr{free} (\tilde{x})$ and $D_{xx}^\mr{free} (\tilde{x})$, and in the presence of a perfect conductor surface, considering the constant distance $\tilde{z} = 0.2$ from the surface, denoted by $D_{zz}^\mr{per} (\tilde{x}, \tilde{z}=0.2)$ and $D_{xx}^\mr{per} (\tilde{x}, \tilde{z}=0.2)$. It illustrates how the presence of the surface can assist with maintaining the relative decay close to zero when the dipoles are oriented parallel to the surface in the $xx$ configuration. In the inset of Fig.~\ref{fig:DD_ff} (a), we plot the time evolution of the concurrence between two emitters at the specific point $(\tilde{x} = 1, \tilde{z} = 0.2)$, marked with the vertical dashed line in Fig.~\ref{fig:DD_ff} (a), calculated based on the master equation (\ref{eq: MasterEq}). The emitters are assumed to be silicon-vacancy (SiV) centers embedded in a nanodiamond, placed near a surface with a transition wavelength $\lambda = 2\pi/k_0 = 737$ nm, and the spontaneous emission rate can be approximated by $\Gamma_0 \approx 1.81 \times 10^{7}~s^{-1}$ \cite{NVSi01}.
 It indicates that the generated entanglement for the $xx$ configuration, with $D_{xx}^{\text{per}} \approx 0$ (solid red curve), is preserved for a long time, while for the $zz$ configuration, with $D_{zz}^{\text{per}} > 0$ (dash-dotted blue curve), it decays to zero significantly faster.
 
The interaction between emitters and the surface can be elucidated by the dipole-image model, where the perfect conductor surface acts as a mirror for the dipoles \cite{BookDispersion01}, as depicted in Fig.~\ref{fig:DD_ff}(b-c).

\begin{figure*}[ht]
    \centering
    \subfloat[]{{\includegraphics[width=0.30\textwidth]{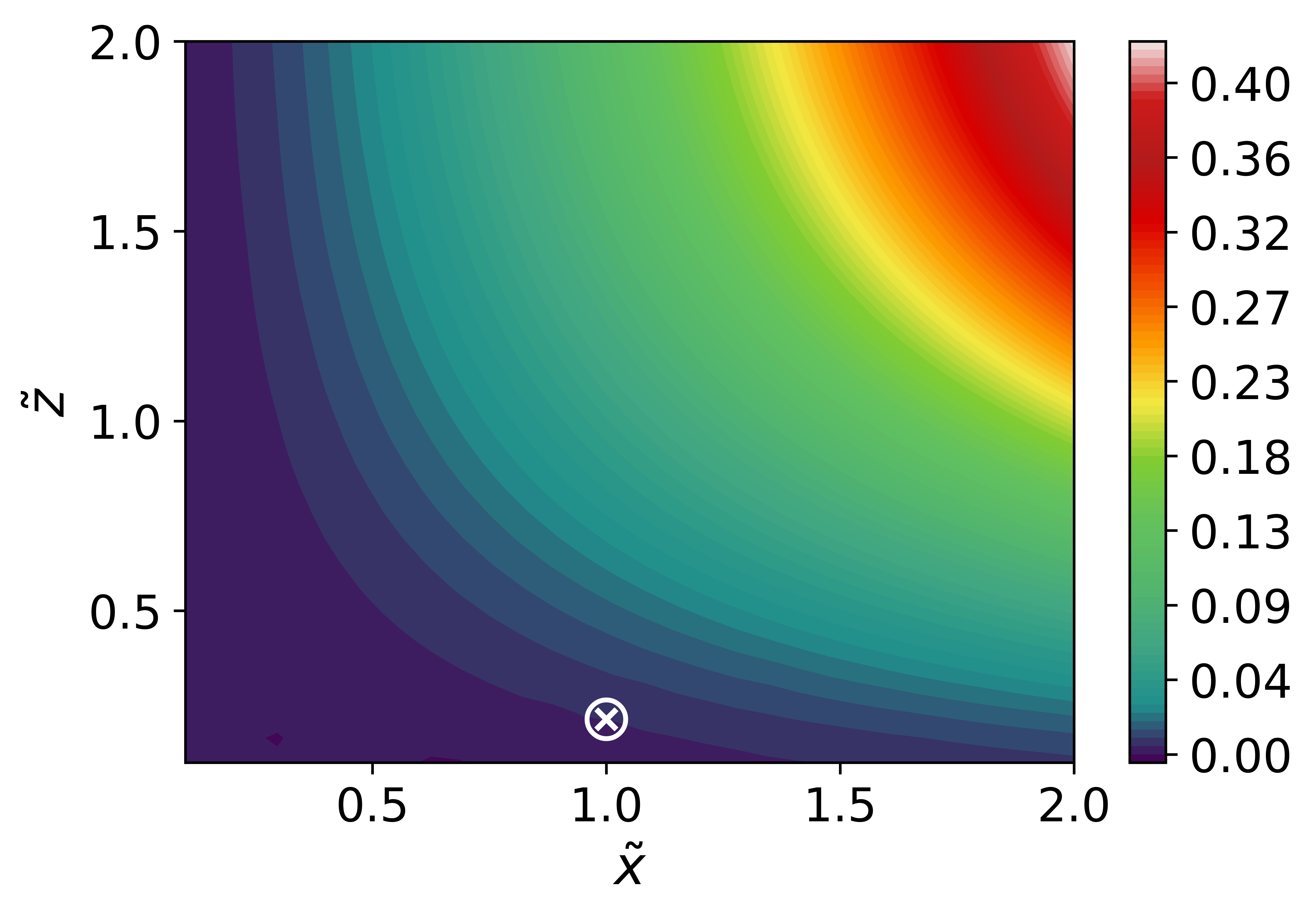}}} \hspace{-0.2cm}
    \subfloat[]{{\includegraphics[width=0.30\textwidth]{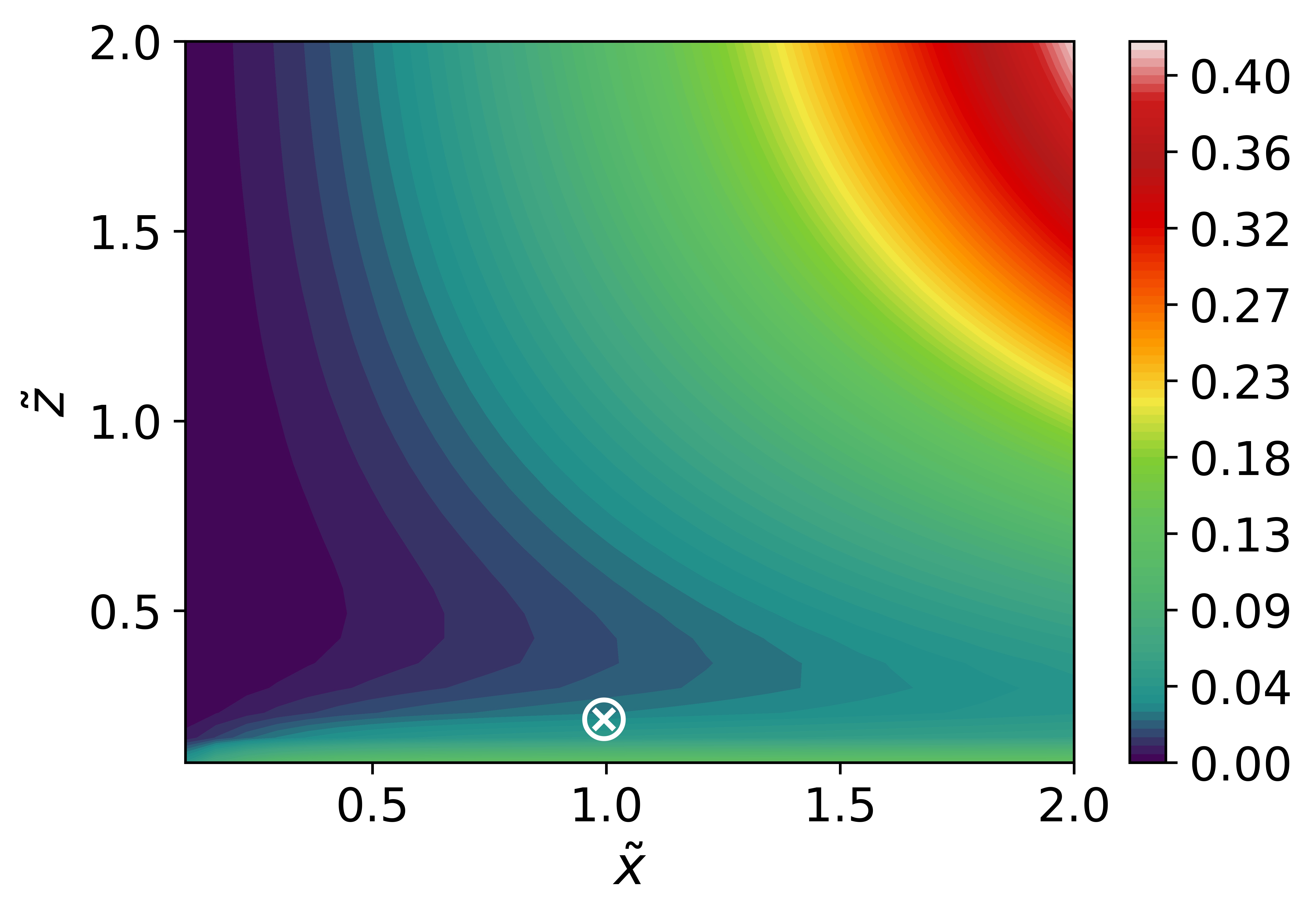}}} \hspace{-0.2cm}
    \subfloat[]{{\includegraphics[width=0.34\textwidth]{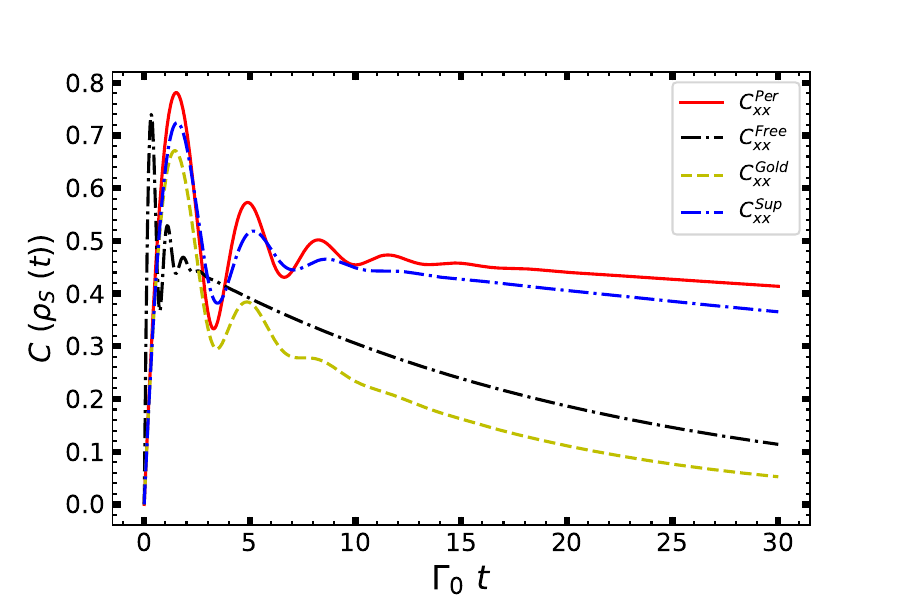}}} \hspace{-0.2cm}
    \caption{Contour plot depicting $D(\tilde{x},\tilde{z})$ in the vicinity of (a) superconducting, and (b) a gold surface where Zero relative decay is required for steady-state entanglement generation. (c) The evolution of concurrence between atoms over time for the $xx$ configuration with $\Tilde{x} = 1$ in free space (black dotted), and at a distance $\Tilde{z} = 0.2$ from the perfect conductor (solid red), the superconducting (blue dash-dotted), and the gold surface (yellow dashed). However, atom-atom entanglement can persist for a long time even if the relative decay does not vanish completely. The selected position ($\Tilde{x}=1,\Tilde{z}=0.2$) is indicated by the white $\otimes$ sign in contour plots.}
    \label{fig:DD_Contour}
\end{figure*}


In the case of  $zz$ configuration, where dipoles are perpendicular to the surface, dipoles' field and their image reinforce each other. This reinforcement leads to a decay rate that approaches $2\Gamma_0$. Conversely, in the $xx$ configuration, where dipoles are oriented parallel to the surface, the dipole and its image fields tend to cancel each other out. Consequently, the emitters experience a nearly zero-field environment, resulting in minimal decay. 
In addition to the dipole-dipole interaction, each emitter is influenced by the other's image. These interactions result in a condition where the $xx$ dipole configuration experiences almost no decay, highlighting how the surface effectively maintains the steady-state entanglement condition by minimizing the relative decay. 

Focusing on $xx$ configuration, we extend our analysis to a metal surface described by the Drude model and a superconducting surface. In this representation, we model the metal surface as gold, setting the plasma frequency to $\omega_p \approx 1.37 \times 10^{16}$ Hz and the loss parameter to $\gamma \approx 5.31 \times 10^{13}$ Hz \cite{GoldRef} where the permittivity described as $\epsilon_{d}(\omega) = 1 - \omega_p^2/(\omega^2 + i\omega \gamma)$~\cite{BookJackson}. Moreover, the permittivity for a London superconductor is given by \cite{Superconducting001}
\begin{align}
    \epsilon_{s}(\omega) = 1 - \frac{c^2}{\omega^2~\lambda_L^2 (T)} + i~\frac{2~c^2}{\omega^2~\delta_L^2 (T)}
\end{align} \label{eq:ep_Sup}

The (square of the) London penetration length and the skin depth are defined as $\lambda_L^2 (T) = \lambda_L^2(0)/[1 - (T/T_c)^4]$, and $\delta_L^2(T) = 2/(\omega \mu_0 \sigma (T/T_c)^4)$, where $\sigma = 2 \times 10^9~\Omega^{-1}$ represents the electrical conductivity. We consider a superconducting Niobium with the critical temperature $T_c = 8.31$~K and $\lambda_L(0) = 35$~nm at a finite temperature $T = 0.01 T_c$.

Fig.~\ref{fig:DD_Contour} (a-b) displays contour plots of $D(x, z)$ near (a) a Niobium superconductor, and (c) a gold surface, as functions of $\tilde{x}$ and $\tilde{z}$. The transition from dark blue to red indicates an increase in relative decay from zero, with dark blue areas highlighting the parameter space where the steady-state or near steady-state entanglement condition is satisfied. According to Fig.~\ref{fig:DD_Contour}(a), the relative decay tends to zero as the distance $\tilde{z}$ from the surface decreases.
For the superconducting surface in the non-retarded regime, $D_{xx}^{\text{Sup}} \propto (T/T_c)^4$ (See \ref{App.C}). Therefore, at low temperatures $T \ll T_c$, the relative decay is similar to that of a perfect conductor. 
In contrast, Fig.~\ref{fig:DD_Contour}(b) shows that the relative decay increases as $\tilde{z}$ decreases near the gold surface. In the non-retarded regime ($\tilde z \ll1$), it can be shown that $D_{xx}^{\text{Gold, sc}} (\tilde{x}, \tilde{z}) \propto \gamma \Gamma_0 / \tilde{z}^3$. This demonstrates that in metals with non-zero loss, the emitters experience a noisier environment in the non-retarded regime as the distance from the metal surface decreases due to significant surface scattering. As illustrated in Fig.~\ref{fig:DD_Contour}(b), an optimal distance from the surface can be determined to minimize the relative decay and maximize surface-mediated entanglement.

After analytically demonstrating the necessity of zero relative decay for steady-state entanglement generation, we numerically identified potential geometric configurations satisfying this condition in Fig.~\ref{fig:DD_Contour} (a)-(b). However, achieving exact zero relative decay may not be feasible in practice. Nonetheless, a significant amount of entanglement could still be maintained for an extended period between the emitters even if the relative decay does not vanish completely. According to our calculations, the relative decay of the system near a perfect conductor behaves similarly to that of the system near a superconducting surface (See \ref{App.C}). The relative decay for the superconducting surface (Fig.~\ref{fig:DD_Contour} (a)) can approach zero at $\tilde{z} \leq 0.2$, resulting in nearly steady-state entanglement, which can be further enhanced as $\tilde{z}$ decreases. On the other hand, near the metal surface, the generated entanglement at $\tilde{z} < 0.2$ decays quickly due to near-field interactions. However, there is an optimal distance that minimizes the relative decay. For instance, the minimum value of the relative decay at $\tilde{x} = 1$ occurs at a distance of approximately $\tilde{z} \approx 0.4$ from the gold surface, resulting in maximum entanglement (Fig.~\ref{fig:DD_Contour} (b)). To compare different surfaces, we consider $\tilde{x} = 1$ ($x \approx 117$ nm) and $\tilde{z} = 0.2$ ($z \approx 23$ nm) to examine the entanglement generation. The selected coordinate is marked in the contour plots (Fig.~\ref{fig:DD_Contour}~(a)-(b)). Furthermore, we consider a system of two emitters in free space at a distance of $\tilde{x} = 1$ without the presence of a surface, which serves as a baseline for free space dipole-dipole entanglement generation. This allows us to separate and elucidate the effect of each surface.

At this point, the relative decay for each type of the surface is obtained as $D_{xx}^{\text{per}}(1,0.2) = 2.2 \times 10^{-3}$, $D_{xx}^{\text{Sup}}(1,0.2) = 4.7 \times 10^{-3}$, and $D_{xx}^{\text{Gold}}(1,0.2) = 0.037$. The time evolution of the concurrence $(C(\rho_S(t)))$ between two emitters at this point, calculated based on the master equation (\ref{eq: MasterEq}), is plotted in Fig.~\ref{fig:DD_Contour} (c). 
The calculations reveal that at scaled time $\Gamma_0 t = 30$, the generated entanglement near the perfect conductor and superconducting surfaces is approximately $C^{\text{per}}_{xx} \approx 0.41$ and $C^{Sup}_{xx} \approx 0.37$, respectively, while it decays to $C^{\text{Gold}}_{xx} \approx 0.05$ near the gold surface, and for free space $C_{xx}^{\text{free}} \approx 0.1$. According to Fig.~\ref{fig:DD_Contour} (c), the concurrence is enhanced by the presence of the perfect conductor (solid red curve) and superconducting surface (dashed-dotted blue curve) compared to free-space (dotted black curve), while the gold surface (dashed yellow curve) reduces entanglement generation. However, it can be shown that at point $\tilde{z} = 0.4$, as an optimum distance from the surface, with the same $\tilde{x} = 1$ the gold surface can also contribute positively, resulting in $C^{\text{Gold}}_{xx} \approx 0.24 > C_{xx}^{\text{free}}$.
\section{Conclusion}
In conclusion, this letter introduced a method for generating steady-state entanglement by investigating the dynamics of a system with two emitters near a surface, considering the CP interaction. We analyzed the emergence of steady-state entanglement through an analytical solution of the master equation, emphasizing the necessary conditions for its realization. We introduced relative decay function $D(\mathbf{r}_1, \mathbf{r}_2) = \Gamma - \Gamma_{12}$ defined by spontaneous emission rates ($\Gamma \equiv \Gamma_{11} = \Gamma_{22}$) and surface-mediated dipole-dipole dissipative interaction ($\Gamma_{12} = \Gamma_{21}$). We demonstrated that achieving an entangled steady-state with a concurrence of $C(\rho_{\text{steady}}) = 0.5$ is possible when the relative decay, expressed as a function of the spatial positions of the emitters, their dipole orientations, and surface properties, equals zero ($D(\mathbf{r}_1, \mathbf{r}_2) = 0$). First, we revealed that to satisfy this condition in a surface-mediated process, the emitters' dipoles must be oriented parallel to the surface. 
Moreover, the surface's optical properties affect the relative decay as a function of the emitters' position. We demonstrated that the superconducting surface closely corresponds to the perfect conductor, maintaining the relative decay near zero. In contrast, due to the metal's loss parameter, the relative decay increases near a metal surface in the non-retarded limit, leading to a faster decay of the generated entanglement. However, we identified an optimal distance from the metal surface that enhances entanglement generation compared to free space and preserves it for a longer time. Furthermore, the presented method can be explored for various surface curvatures~\cite{ScheelMain,CurvaturePRA01,NanoFiber01}, leading to different geometries, to identify the optimal conditions for generating and preserving entanglement.
The proposed method hints at potential applications in near-surface quantum metrology and sensing, where entanglement could play a role in enhancing precision measurements and sensing capabilities. Furthermore, entanglement generation and decoherence from quantum fluctuations are a subject of broader interest in QED and macroscopic quantum systems\cite{Anupam01,Anupam02,Velatko01,CP-coh001,CP-coh002,CP-coh003,Con01,Con02}. 
\subsection{Acknowledgments}
We gratefully acknowledge Stefan Scheel for bringing Ref. [36] to our attention after the completion of this work.  This work was supported by the Scientific and Technological Research Council of T\"{u}rkiye (T\"{U}B\.{I}TAK) under Project Numbers 120F089 and 123F150. M. I., O. P., and \"{O}. M. thank T\"{U}B\.{I}TAK for their support. K.S. acknowledges support from the National Science Foundation under Award No. PHY-2418249, by the John Templeton Foundation under Award No. 62422, and Army Research Office under Award No. W911NF2410080.

\appendix

\section{Surface-Mediated vacuum ElectroMagnetic Field}\label{App.A}
The Hamiltonian for the vacuum EM field near a surface in macroscopic QED formalism is written as \cite{BookDispersion01}
\begin{align}
    \hat{H}_F = \sum_{\lambda=e,m} \int d^3\textbf{r}' \int d\omega ~\hbar \omega ~ \hat{f}_{\lambda}^{\dagger} (\textbf{r}, \omega).\hat{f}_{\lambda} (\textbf{r}, \omega).
\end{align}
Here, $\hat{f}_{\lambda} (\mathbf{r}, \omega)$ and $\hat{f}_{\lambda}^{\dagger} (\mathbf{r}, \omega)$ are the bosonic annihilation and creation operators in the presence of the medium, 
as defined in the main text. They follow the canonical commutation relations $[\hat{f}_{\lambda} (\mathbf{r}, \omega), \hat{f}_{\lambda}^{\dagger} (\mathbf{r}, \omega)] = \delta_{\lambda \lambda^{'}} \delta (\mathbf{r} - \mathbf{r}^{'}) \delta (\omega - \omega^{'})$.

The electric field operator at position $\mathbf{r}$ is given by Eq. (1) in the main text,
wherein the coefficients $\overline{\mathbf{G}}_{\lambda}(\textbf{r},\textbf{r}',\omega)$ defined by
\begin{align}
    \overline{\mathbf{G}}_{e}(\textbf{r},\textbf{r}',\omega) = i \frac{\omega^2}{c^2}\sqrt{\frac{\hbar}{\pi \epsilon_0}Im[\epsilon(r^{'},\omega)]}~\overline{\mathbf{G}}(\textbf{r},\textbf{r}',\omega), \nonumber \\
     \overline{\mathbf{G}}_{m}(\textbf{r},\textbf{r}',\omega) = i \frac{\omega^2}{c^2}\sqrt{\frac{\hbar}{\pi \epsilon_0}\frac{Im[\mu(r^{'},\omega)]}{|\mu(r^{'},\omega)|^2}}~\nabla \times \overline{\mathbf{G}}(\textbf{r},\textbf{r}',\omega).
\end{align}
where the Green's tensor satisfies the inhomogeneous Helmholtz equation in the presence of the medium as follow
\begin{align}\label{Eq:Helmoltz}
    \mathbf{\nabla}\times\mathbf{\nabla}\times\Bar{G}(\mb{r}, \mb{r}', \omega) - \epsilon(\mathbf{r},\omega)\frac{\omega^2}{c^2} \Bar{G}(\mb{r}, \mb{r}', \omega) = \delta (\mb{r} - \mb{r}'), 
\end{align}
where $\bar{G}(\mathbf{r}, \mathbf{r}', \omega) = \bar{G}_{\text{free}}(\mathbf{r}, \mathbf{r}', \omega) + \bar{G}_{\text{sc}}(\mathbf{r}, \mathbf{r}', \omega)$ incorporates the free space and scattering components of the total Green's tensor. The permittivity and permeability of the medium are denoted by $\epsilon(\mathbf{r},\omega)$, and $\mu(\mathbf{r},\omega)$, respectively.
\par Considering a dipole near an infinite half-space surface located at point $\mathbf{r}_1$, the scattering Green's tensor is obtained as:
\begin{align}
    \Bar{G}_{\text{sc}}(x,z,\omega) &= \frac{i}{8 \pi} \int_0^{\infty} dk_{\parallel} \frac{k_{\parallel}}{k_{\bot}} e^{2 i k_{\bot} z} \biggl[ \begin{pmatrix}
J_0(k_{\parallel} x) + J_2 (k_{\parallel} x) & 0 & 0\\
0 & J_0(k_{\parallel} x) - J_2 (k_{\parallel} x) & 0\\
0 & 0 & 0
\end{pmatrix} r_s \nonumber \\
&+ \frac{c^2}{\omega^2} \begin{pmatrix}
-k_{\bot}^2 [J_0(k_{\parallel} x) - J_2 (k_{\parallel} x)] & 0 & -2 i k_{\parallel} k_{\bot} J_0(k_{\parallel} x)\\
0 & -k_{\bot}^2 [J_0(k_{\parallel} x) + J_2 (k_{\parallel} x)] & 0\\
2 i k_{\parallel} k_{\bot} J_0(k_{\parallel} x) & 0 & 2 k_{\parallel}^2 J_0(k_{\parallel} x)
\end{pmatrix} r_p \biggl]
\end{align}

Here, we consider$|\mathbf{r}_1 - \mathbf{r}_2| = x$ and $(\mathbf{r}_1 + \mathbf{r}_2) \cdot \hat{z} = z $. The Fresnel coefficients for $s$- and $p$-polarized light reflecting from the surface are denoted as $r_s= \frac{k_{\bot} - k_{\bot}^1}{k_{\bot} + k_{\bot}^1}$ and $r_p= \frac{\epsilon(\omega) k_{\bot} - k_{\bot}^1}{\epsilon(\omega) k_{\bot} + k_{\bot}^1}$, respectively, where  $k_{\perp}^2 = \omega^2 - k_{\parallel}^2$, 
where $k^1_{\bot} = \sqrt{\epsilon(\omega) \omega^2 / c^2 + k_{\parallel}^2}$.
\par Moreover, the free Green's tensor between $\mathbf{r}_1$, and $\mathbf{r}_2$ gives 
\begin{align}
    \Bar{G}_{\text{free}}(\mathbf{r_1},\mathbf{r_2},\omega) = -\frac{c^2 e^{i \omega x/c}}{4 \pi \omega^2 x^3}  \begin{pmatrix}
g(-i \omega x/c) - h(-i \omega x /c) & 0 & 0\\
0 & g(-i \omega x/c) & 0\\
0 & 0 & g(-i \omega x/c)
\end{pmatrix} 
\end{align}


\section{Steady-State Calculations} \label{App.B}
The equations of the motion for the density matrix elements can be obtained from the master equation (Eq. 2 in the main text) as follows:
\begin{subequations}
\renewcommand{\theequation}{\theparentequation.\arabic{equation}}
\begin{align}
   \frac{d\rho_{11}(t)}{dt} &= -2\Gamma~\rho_{11}(t) \\
   \frac{d\rho_{44}(t)}{dt} &= \Gamma~\bigg(\rho_{22}(t) + \rho_{33}(t)\bigg) + \Gamma_{12}~\bigg(\rho_{23}(t) + \rho_{32}(t)\bigg) \\ 
   \frac{d\rho_{22}(t)}{dt} &= -\frac{i}{\hbar} \Omega_{12}\bigg(\rho_{32}(t) - \rho_{23}(t)\bigg) + \Gamma~\bigg(\rho_{11}(t) - \rho_{22}(t)\bigg) - \frac{\Gamma_{12}}{2}~\bigg(\rho_{23}(t) + \rho_{32}(t)\bigg) \\ 
   \frac{d\rho_{33}(t)}{dt} &= -\frac{i}{\hbar} \Omega_{12}\bigg(\rho_{23}(t) - \rho_{32}(t)\bigg) + \Gamma~\bigg(\rho_{11}(t) - \rho_{33}(t)\bigg) - \frac{\Gamma_{12}}{2}~\bigg(\rho_{23}(t) + \rho_{32}(t)\bigg) \\
   \frac{d\rho_{23}(t)}{dt} &= -\frac{i}{\hbar} \Omega_{12}\bigg(\rho_{33}(t) - \rho_{22}(t)\bigg) - \Gamma~\rho_{23}(t) + \frac{\Gamma_{12}}{2}~\bigg(2\rho_{11}(t) - \rho_{22}(t)- \rho_{33}(t)\bigg) \\
   \frac{d\rho_{32}(t)}{dt} &= -\frac{i}{\hbar} \Omega_{12}\bigg(\rho_{22}(t) - \rho_{33}(t)\bigg) - \Gamma~\rho_{32}(t) + \frac{\Gamma_{12}}{2}~\bigg(2\rho_{11}(t) - \rho_{22}(t)- \rho_{33}(t)\bigg) 
\end{align}
\end{subequations}
Considering initial state $\ket{\psi_0 (t=0)} = \ket{eg}$, the solution of the elements of the density matrix is obtained as shown in Eqs.~(8.1)-(8.6) in the main text where the steady state can be obtained considering $\Gamma = \Gamma_{12}$ in the limit $t \xrightarrow{} \infty$ as

\begin{align}
    \rho_{steady} &= \frac{1}{2}\ket{gg}\bra{gg} + \frac{1}{4}\bigg[ \bigg(\ket{eg}\bra{eg} + \ket{ge}\bra{ge}\bigg) -  \bigg(\ket{eg}\bra{ge} + \ket{ge}\bra{eg}\bigg) \bigg] \nonumber \\
    & = \frac{1}{2}\bigg(\ket{gg}\bra{gg} + \ket{\psi_{sub}}\bra{\psi_{sub}}\bigg)
\end{align}
Here, $\ket{\psi_{sub}} = (\ket{eg} - \ket{ge})/\sqrt{2}$.


\subsection{Concurrence}
To quantify the quantum entanglement between two atoms in the resulting steady-state, the concurrence is defined as $C(\rho) = \mr{max}\cbkt {0,\lambda_1 - \lambda_2 - \lambda_3 - \lambda_4}$ \cite{concurrence01,concurrence02,ConcurrenceRef}, where  $\lambda_i$ for $i = 1, 2, 3, 4$ are the eigenvalues, arranged in decreasing order, of the Hermitian matrix $R = \sqrt{\sqrt{\rho}\tilde{\rho}\sqrt{\rho}}$, with $\tilde{\rho} = (\sigma_y \otimes \sigma_y)\rho^{*}(\sigma_y \otimes \sigma_y)$. Accordingly, the concurrence of the resulting steady-state, Eq. (8) in the main text, becomes $C(\rho_{\text{steady}}) = 0.5.$
\section{Relative Decay}\label{App.C}
Defining the relative decay as $D=\Gamma - \Gamma_{12}$, we examine the behavior of the dissipative coefficients in two dipole configurations, $xx$ and $zz$. The dissipative coefficients are defined as \(\Gamma = \Gamma_0 + \Gamma^{sc}\) and \(\Gamma_{12} = \Gamma_{12}^{free} + \Gamma_{12}^{sc}\), where each includes both the free-space and surface-mediated components. Here, $\Gamma = \Gamma_0 + \Gamma^{sc}$ represents the decoherence effect due to the single atom's spontaneous emission and the scattering contribution, given as:

\small
\begin{align}
    \Gamma^{zz} (\tilde{z})& = \Gamma_0 + \Gamma^{zz,sc} (\tilde{z}) = \Gamma_0 + \frac{3\Gamma_0}{8\tilde{z}^3}\biggl(\bigl(\sin(2\tilde{z}) - 2\tilde{z}\cos(2\tilde{z})\bigl)~\re[r_p] + \bigl(2\tilde{z}\sin(2\tilde{z}) + \cos(2\tilde{z})\bigl)~\im[r_p]\biggl) \nonumber \\
    \Gamma^{xx} (\tilde{z})& =\Gamma_0 + \Gamma^{xx,sc} (\tilde{z}) \nonumber\\
    &=\Gamma_0 - \frac{3\Gamma_0}{16\tilde{z}^3}\biggl(\bigl(2\tilde{z}\cos(2\tilde{z}) + (2\tilde{z}^2 - 1)\sin(2\tilde{z})\bigl)~\re[r_p] - \bigl(2\tilde{z}\sin(2\tilde{z}) + (1-2\tilde{z}^2)\cos(2\tilde{z})\bigl)~\im[r_p]\biggl) 
\end{align}
\normalsize
\par Additionally, \(\Gamma_{12}\) characterizes the cooperative dissipation. This term plays a key role in the system's coherent effects, as the two atoms are weakly coupled to the same bath under the Markov approximation. The free-space contribution \(\Gamma_{12}^{free}\) is a function of the interatomic distance $x$, while the scattering component \(\Gamma_{12}^{sc}\) depends on \(x\), \(z\), and the medium's properties. $\Gamma_{12}$ for two  dipole configurations are calculated as: 
\small
\begin{align}
     \Gamma_{12}^{zz}(\tilde{x}, \tilde{z}) &= \frac{2\mu_0 \omega_0^2 |d|^2}{\hbar} \im[\mathbf{z}\cdot (\Bar{G}_{{\text{free}}}(\mathbf{r}_1,\mathbf{r}_2,\omega_0)+ \Bar{G}_{{\text{sc}}}(\mathbf{r}_1,\mathbf{r}_2,\omega_0) ) \cdot \mathbf{z}]\nonumber \\
    & = \frac{3\Gamma_0}{2 \tilde{x}^3} \left((1 - \tilde{x}^2)\sin(\tilde{x}) - \tilde{x}\cos(\tilde{x})\right) \nonumber \\
    &+ \frac{3\Gamma_0}{2} \biggl(\int_{0}^{1} d\tilde{k}_{\bot} (1 - \tilde{k}_{\bot}^2) J_0(\tilde{x}\sqrt{1 - \tilde{k}_{\bot}^2}) \left(\cos(2 \tilde{k}_{\bot}\tilde{z})~\re[r_p] - \sin(2 \tilde{k}_{\bot}\tilde{z}])~\im[r_p]\right) \nonumber\\ 
    &+ \int_{0}^{\infty} d\tilde{\kappa}_{\bot} e^{- 2 \tilde{\kappa}_{\bot}\tilde{z}}~(1 + \tilde{\kappa}_{\bot}^2) J_0(\tilde{x}\sqrt{1 + \tilde{\kappa}_{\bot}^2})~\im[r_p]\biggl), \nonumber \\
    \Gamma_{12}^{xx}(\tilde{x}, \tilde{z}) &= \frac{2\mu_0 \omega_0^2 |d|^2}{\hbar} \im[\mathbf{x}\cdot (\Bar{G}_{{\text{free}}}(\mathbf{r}_1,\mathbf{r}_2,\omega_0) + \Bar{G}_{{\text{sc}}}(\mathbf{r}_1,\mathbf{r}_2,\omega_0) ) \cdot \mathbf{x}]\nonumber \\
    & = - \frac{3\Gamma_0}{ \tilde{x}^3} \left(\sin(\tilde{x}) - \tilde{x}\cos(\tilde{x})\right) \nonumber \\
    &- \frac{3\Gamma_0}{ \tilde{x}^3} \left(\sin(\tilde{x}) - \tilde{x}\cos(\tilde{x})\right) -\frac{3\Gamma_0}{4} \biggl(\int_{0}^{1} d\tilde{k}_{\bot} \tilde{k}_{\bot}^2 (J_0(\tilde{x}\sqrt{1 - \tilde{k}_{\bot}^2}) - J_2(\tilde{x}\sqrt{1 - \tilde{k}_{\bot}^2})) \nonumber \\
    &\times \left(\cos(2 \tilde{k}_{\bot}\tilde{z})~\re[r_p] - \sin(2 \tilde{k}_{\bot}\tilde{z}])~\im[r_p]\right) \nonumber \\
    &- \int_{0}^{\infty} d\tilde{\kappa}_{\bot} e^{- 2 \tilde{\kappa}_{\bot}\tilde{z}}~ \tilde{\kappa}_{\bot}^2 (J_0(\tilde{x}\sqrt{1 + \tilde{\kappa}_{\bot}^2}) - J_2(\tilde{x}\sqrt{1 + \tilde{\kappa}_{\bot}^2}))~\im[r_p]\biggl)
\end{align}
\normalsize

In the following, we describe three different surfaces with different permittivities.
\subsection{Perfect Conductor}
\begin{figure}[t]
    \centering
  \subfloat[]{{\includegraphics[width=0.45\textwidth]{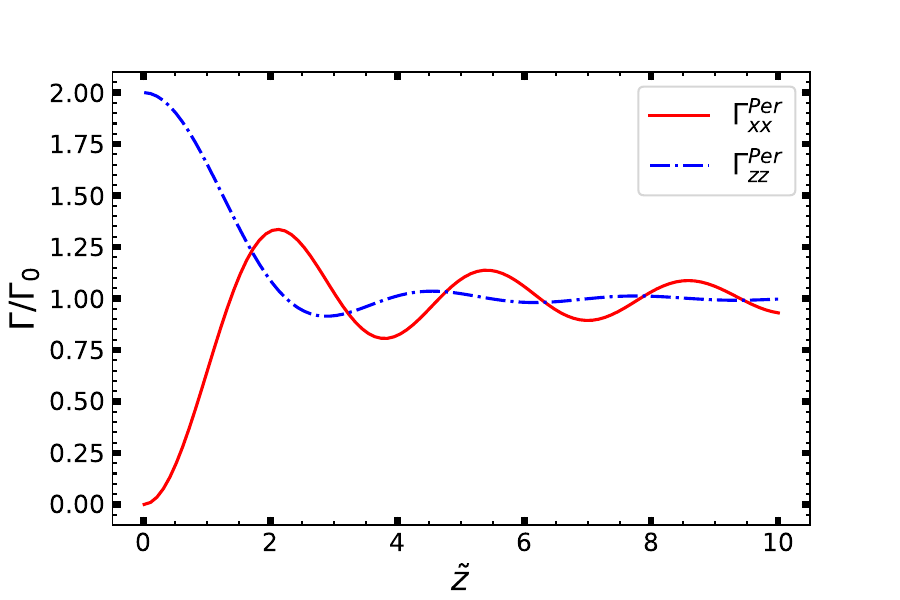} }} 
    \subfloat[]{{\includegraphics[width=0.45\textwidth]{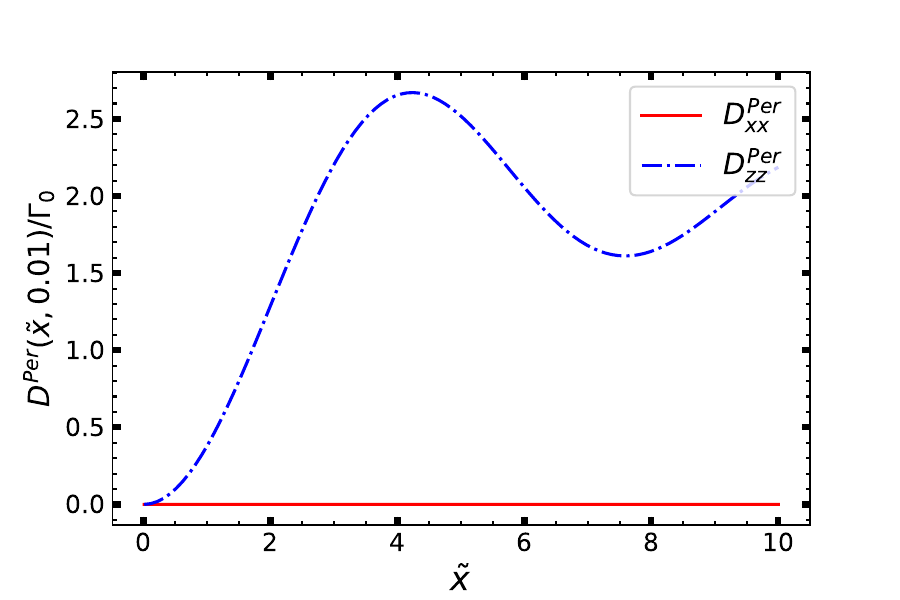} }}
     \caption{(a) The single emitter decay $\Gamma(\tilde{z})$ as a function of $\tilde{z}$ and (b) The relative decay $D(\tilde{x},\tilde{z})$ as a function of $\tilde{x}$ where $\tilde{z} = 0.01$ for $xx$ (the solid red curve), and $zz$ (dashed blue curve) configurations near a perfect conductor. }
    \label{fig:Decay_per}
\end{figure}

In a perfect conductor $\epsilon (\omega) \xrightarrow{} \infty$, and as a results the Fresnel coefficients  become
\begin{align}
    r_s^{Per} &= -1 \nonumber \\
    r_p^{Per} &= 1
\end{align}
Accordingly, the scattering parts are obtained
\begin{align}\label{eq:Gsczz}
    \Gamma_{11}^{zz} (\tilde{z})& = \frac{3\Gamma_0}{8\tilde{z}^3}\bigl(\sin(2\tilde{z}) - 2\tilde{z}\cos(2\tilde{z})\bigl) 
\end{align}
\begin{align}\label{eq:Gsc12zz}
    \Gamma_{12}^{\text{sc},zz} (\tilde{x},\tilde{z}) &= \frac{3\Gamma_0}{2} \biggl(\int_{0}^{1} d\tilde{k}_{\bot} (1 - \tilde{k}_{\bot}^2) J_0(\tilde{x}\sqrt{1 - \tilde{k}_{\bot}^2}) \cos(2 \tilde{k}_{\bot}\tilde{z}) \biggl),
\end{align}
\begin{align}\label{eq:Gscxx}
    \Gamma_{11}^{sc,xx} (\tilde{z})& = -\frac{3\Gamma_0}{16\tilde{z}^3}\bigl(2\tilde{z}\cos(2\tilde{z}) + (2\tilde{z}^2 - 1)\sin(2\tilde{z})\bigl)
\end{align}

\begin{align}\label{eq:Gsc12xx}
    \Gamma_{12}^{\text{sc},xx} (\tilde{x},\tilde{z}) &= -\frac{3\Gamma_0}{4} \biggl(\int_{0}^{1} d\tilde{k}_{\bot} \tilde{k}_{\bot}^2 (J_0(\tilde{x}\sqrt{1 - \tilde{k}_{\bot}^2}) - J_2(\tilde{x}\sqrt{1 - \tilde{k}_{\bot}^2})) \left(\cos(2 \tilde{k}_{\bot}\tilde{z}) \right)\biggl)
\end{align}
Fig.~\ref{fig:Decay_per} (a) shows $\Gamma$ as a function of $\tilde{z}$ for $zz$ (dashed blue curve), and $xx$ (solid red curve) dipole orientations. According to Fig.~\ref{fig:Decay_per} (a), the competition between \(\Gamma^{sc}\) and \(\Gamma_0\) reduces the decoherence effect in the parallel dipole configuration, thereby helping to preserve the system's coherence. Here, the scattering component \(\Gamma^{sc}\) depends on the distance from the surface (\(z\)) and the surface's material properties.
According to Eqs. (\ref{eq:Gsczz}) and (\ref{eq:Gscxx}), in the non-retarded regime where $\tilde{z}\ll1$, it can be observed that $\Gamma_{11}^{sc,zz}\xrightarrow{}\Gamma_0$, while $\Gamma_{11}^{sc,xx}\xrightarrow{}-\Gamma_0$. Consequently, in the non-retarded limit, $\Gamma \xrightarrow{}2\Gamma_0$ and $\Gamma \xrightarrow{}0$. As $\tilde{z}$ increases, it tends to $\Gamma_0$, which diminishes the surface's effect, as shown in Fig.~\ref{fig:Decay_per} (a). 
The scattering part of the relative decay near a perfect conductor is calculated for each dipole configuration based on Eq.~(7) in the main text as
\begin{align}\label{Eq:DD-sc-perfect}
    D_{zz}^{\text{sc},\mr{per} }(\tilde{x},\tilde{z}) = & \frac{3 \Gamma_0}{2} ~\sbkt{\frac{\sin(2\tilde{z}) - 2\tilde{z} \cos(2\tilde{z})}{4\tilde{z}^3} - \mathcal{G}_{zz} (\tilde{x},\tilde{z})} \non \\ 
    D_{xx}^{\text{sc},\mr{per} }(\tilde{x},\tilde{z}) = & -\frac{3\Gamma_0}{4}\sbkt{\frac{(4\tilde{z}^2 - 1) \sin(2\tilde{z}) + 2\tilde{z}\cos(2\tilde{z})}{4\tilde{z}^3} - \mathcal{G}_{xx} (\tilde{x},\tilde{z})}.
\end{align}

Here we have introduced

\begin{align}
    \mathcal{G}_{zz} (\tilde{x},\tilde{z}) \equiv&  \int_{0}^{1} d\tilde{k}_{\bot} \cos(2 \tilde{k}_{\bot}\tilde{z})~(1 - \tilde{k}_{\bot}^2) J_0(\tilde{k}_{\parallel}\tilde{x}) \nonumber \\
    \mathcal{G}_{xx} (\tilde{x},\tilde{z}) \equiv&  \int_{0}^{1} d\tilde{k}_{\bot} \cos(2 \tilde{k}_{\bot}\tilde{z})\left([J_0(\tilde{k}_{\parallel}\tilde{x}) +J_2(\tilde{k}_{\parallel}\tilde{x})] - \tilde{k}_{\bot}^2 [J_0(\tilde{k}_{\parallel}\tilde{x}) - J_2(\tilde{k}_{\parallel}\tilde{x})]\right),
\end{align}
where $\tilde{k}_{\parallel} = \sqrt{1 - \tilde{k}_{\bot}^2}$. 
\begin{figure}[t]
    \centering
  \subfloat[]{{\includegraphics[width=0.45\textwidth]{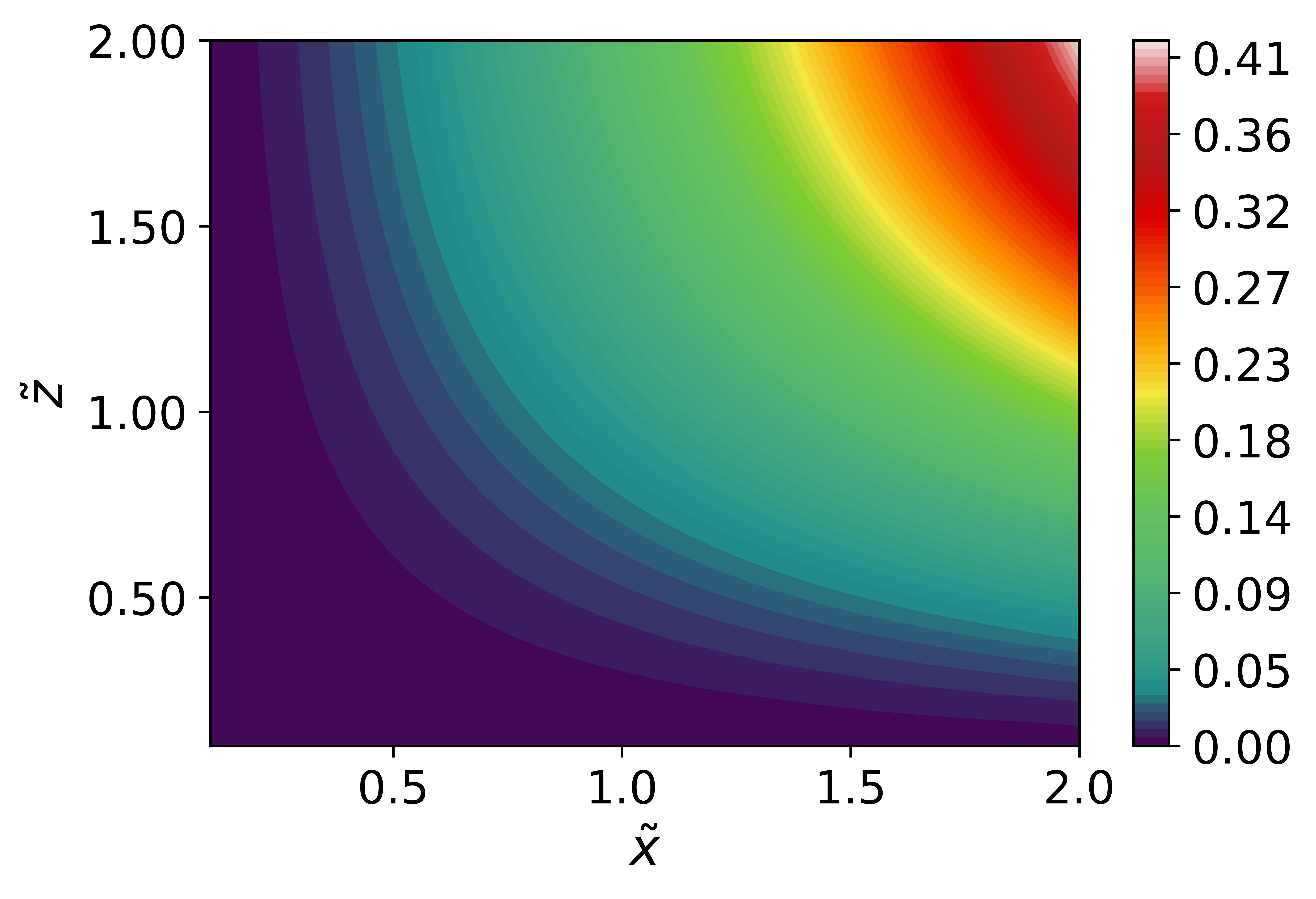} }} 
     \caption{  The contour plot of the relative decay as a function of $\tilde{x}$, and $\tilde{z}$ near a perfect conducting surface for $xx$ configuration. }
    \label{fig:DDPer}
\end{figure}
Furthermore, we calculate the relative decay as a function of $\tilde{x}$ in the non-retarded limit, where the emitters are assumed to be at a constant distance $\tilde{z}$ from the surface, and plot it for two dipole configurations in Fig.~\ref{fig:Decay_per}(b). The figure shows that in the $xx$ configuration, the relative decay remains close to zero, while in the $zz$ configuration it increases as $\tilde{x}$ increases. The contour plot of the relative decay is shown in Fig.~\ref{fig:DDPer} for the $xx$ configuration as a function of $\tilde{x}$ and $\tilde{z}$. 
\par
\begin{figure}[t]
    \centering
  \subfloat[]{{\includegraphics[width=0.45\textwidth]{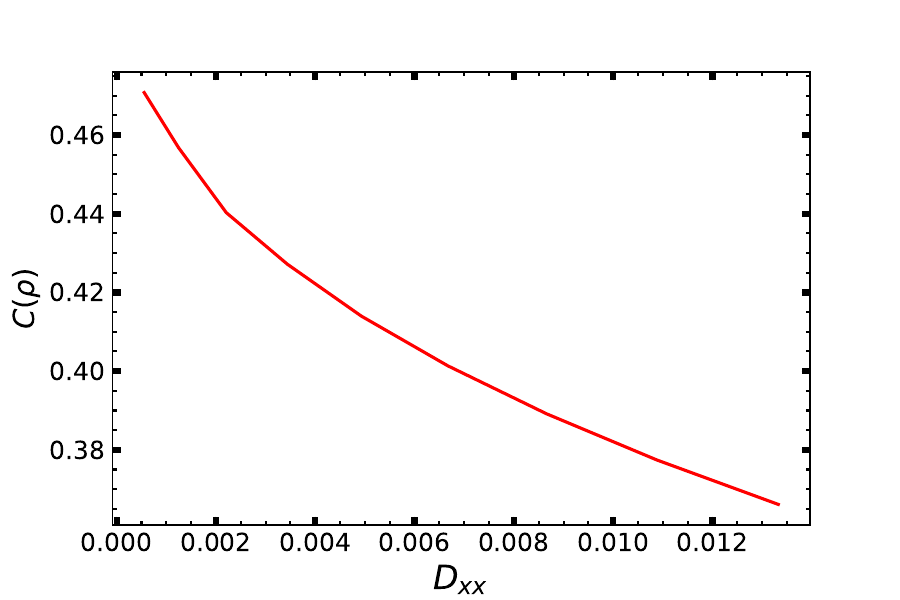} }} 
     \caption{  The concurrence $C(\rho)$ versus the relative decat $D_{xx}$. }
    \label{fig:DC}
\end{figure}

 Moreover, Fig.~\ref{fig:DC} illustrates the relationship between the relative decay \(D_{xx}\) and the concurrence of the system, \(C(\rho)\), showing that \(C(\rho)\) decreases as \(D_{xx}\) increases. For this analysis, we consider the dipole configuration \(xx\), in the optimal condition, and vary \(D_{xx}\) by changing \(\tilde{z}\), while keeping \(\tilde{x} = 1\) fixed. The concurrence is calculated in a scaled time \(\Gamma_0 t = 20\).
\subsection{Drude Model (Metal Surface)}
The Drude model describes the permittivity of a metal surface as follows:
\begin{align}
    \epsilon(\omega) = 1 - \frac{\omega_p^2}{\omega^2 + i\omega~\gamma},
\end{align}
where $\omega_p$ and $\gamma$ describe the plasma frequency and loss parameter of the medium.
\par In non-retarded limit, it can be supposed that $k^1_{\bot} = \sqrt{\epsilon(\omega) \omega^2 / c^2 + k_{\parallel}^2} \approx k_{\bot}$, the Fresnel coefficients can be thus approximated as \cite{BookDispersion01,Collective-Kanu01}:
\begin{figure}[t]
    \centering
  {{\includegraphics[width=0.45\textwidth]{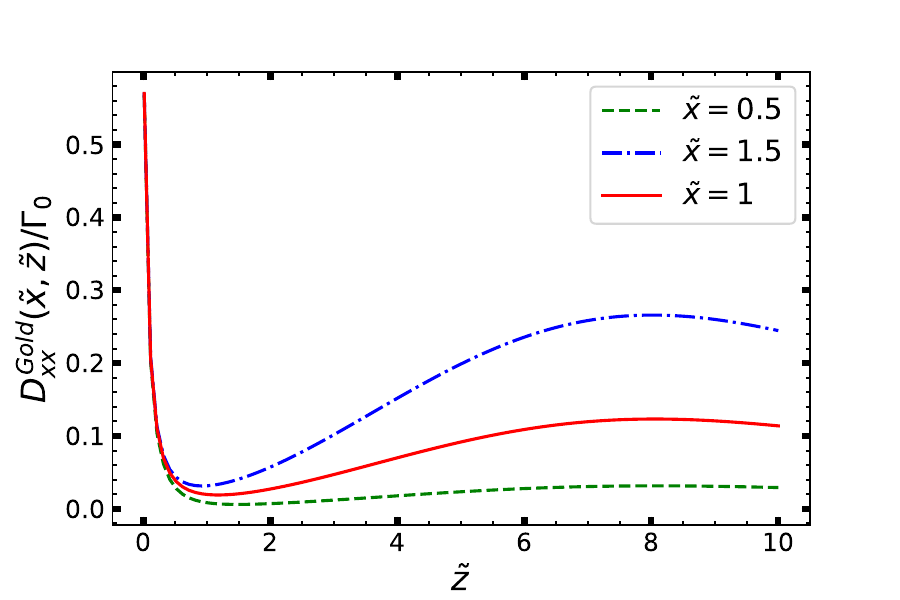} }} 
     \caption{  The relative decay $D(\tilde{x},\tilde{z})$ as a function of $\tilde{x}$ for $\tilde{z} = 0.5, 1, 1.5$ near a gold surface. }
    \label{fig:Decay_drude}
\end{figure}
\begin{align} \label{Eq:FresnelCoff}
    r_s =& \frac{k_{\bot} - k_{\bot}^1}{k_{\bot} + k_{\bot}^1} \approx 0 \nonumber \\
    r_p = &\frac{\epsilon(\omega) k_{\bot} - k_{\bot}^1}{\epsilon(\omega) k_{\bot} + k_{\bot}^1} \approx \frac{\epsilon(\omega) - 1}{\epsilon(\omega) + 1}\approx \frac{-\omega_p^2}{(2\omega^2 - \omega_p^2) + 2i~\omega~\gamma}
\end{align}

As a result, the relative decay for $xx$ configuration is given
\begin{align}
     D_{xx}^{Drude} (\tilde{x},\tilde{z}) =  \left[\left(1 - \frac{3~\left(\sin(\tilde{x}) - \tilde{x}\cos(\tilde{x})\right)}{ \tilde{x}^3} \right) + \frac{3}{8} \left(\frac{~\omega\gamma}{\omega_p^2}\right)\left(\frac{1}{\tilde{z}^3} - 4\mathcal{F}_{xx}(\tilde{x},\tilde{z}) \right)\right]\Gamma_0
\end{align}
Here, we introduced 
\small
\begin{align}
    \mathcal{F}_{xx} (\tilde{x},\tilde{z}) &\equiv \int_{0}^{\infty} d\tilde{\kappa}_{\bot} e^{- 2 \tilde{\kappa}_{\bot}\tilde{z}}~ \tilde{\kappa}_{\bot}^2 (J_0(\tilde{\kappa}_{\parallel}\tilde{x}) - J_2(\tilde{\kappa}_{\parallel}\tilde{x}))
\end{align}
\normalsize
Unlike in a perfect conductor, in the non-retarded limit, $D_{xx}^{\text{Drude}} \propto \gamma\Gamma_0/\tilde{z}^3$, indicating that the relative decay cannot be zero unless considering a lossless metal.
\begin{figure}[t]
    \centering
  {{\includegraphics[width=0.45\textwidth]{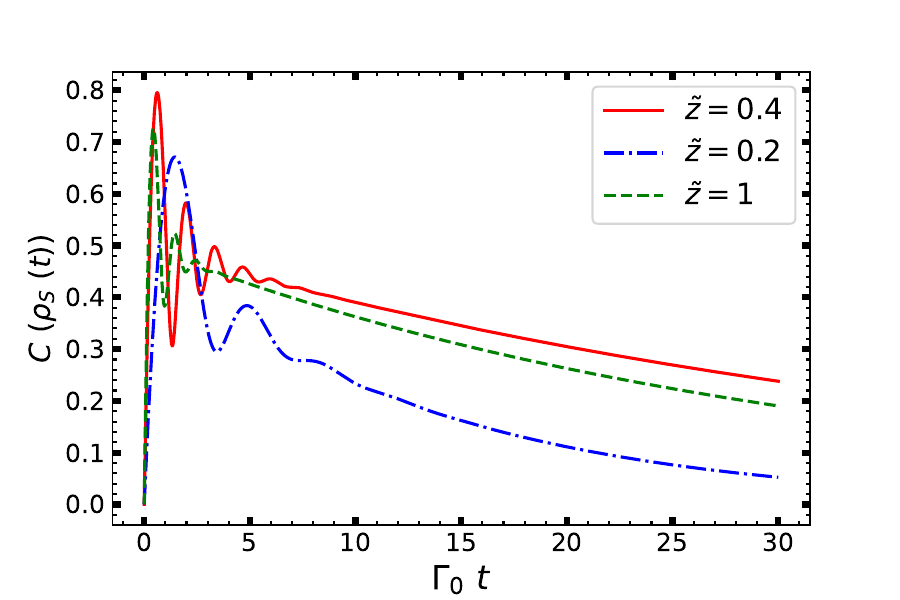} }} 
     \caption{  Evolution of concurrence $C_{xx}^{Gold}(\tilde{x},\tilde{z})$ vs. scaled time $\Gamma_0 t$ for a constant $\tilde{x} = 1$ and at distances $\tilde{z} = 0.2, 0.4, 1$ from the gold surface. }
    \label{fig:Con_drude}
\end{figure}
We plot the relative decay in Fig.~\ref{fig:Decay_drude} for a gold surface using the Drude model, with a plasma frequency of $\omega_p \approx 1.37 \times 10^{16}$ Hz and a loss parameter of $\gamma \approx 5.31 \times 10^{13}$ Hz \cite{GoldRef}. Fig.~\ref{fig:Decay_drude} shows the relative decay as a function of $\tilde{z}$ for three values of $\tilde{x}:$ 0.5, 1, and 1.5. As discussed, the relative decay increases exponentially as $\tilde{z}$ decreases. However, the figure also shows that there is an optimum distance from the surface that minimizes the relative decay, and this optimum distance varies for different values of $\tilde{x}$. In this regard, Fig.~\ref{fig:Con_drude} indicates the time-dependent concurrence of the system for distances $\tilde{z} = 0.2, 0.4, 1$ while keeping $\tilde{x} = 1$ as the constant distance between emitters. As it shows, at $\tilde{z} = 0.4$, where $D_{xx}^{Gold} (\tilde{x} = 1, \tilde{z})$ is minimum, the generated entanglement is maximum, and at time $\Gamma_0 t = 30$ it gives $C_{xx}^{Gold} \approx 0.24$. It thus demonstrates that the medium-mediated interaction enhances entanglement generation at the optimum distance from a metal surface.

\subsection{Superconducting Surface}
In the low-frequency regime, the dielectric function of a London superconductor reads \cite{Superconducting001}
\begin{align}\label{eq:Sup-permittivity}
    \epsilon_{s}(\omega) = 1 - \frac{c^2}{\omega^2~\lambda_L^2 (T)} + i~\frac{2~c^2}{\omega^2~\delta^2 (T)}
\end{align}
Here, the London penetration length and the skin depth are defined as $\lambda_L^2 (T) = \lambda_L^2(0)/[1 - (T/T_c)^4]$, and $\delta_L^2(T) = 2/(\omega \mu_0 \sigma (T/T_c)^4)$, where $\sigma = 2 \times 10^9~\Omega^{-1}$ represents the electrical conductivity. We consider a superconducting Niobium with $T_c = 8.31$~K and $\lambda_L(0) = 35$~nm at a finite temperature $T = 83.1$~mK.
Replacing Eq.(\ref{eq:Sup-permittivity}) into (\ref{Eq:FresnelCoff}), the Fresnel coefficients for a superconducting medium in non-retarded regime are given by
\begin{align}
    r_s &\approx 0 \nonumber \\
    r_p &\approx \frac{\delta_L^2(T) - 2 i \lambda_L^2(T)}{\delta_L^2(T) - 2 i \lambda_L^2(T) -2 \omega^2 \delta_L^2(T) \lambda_L^2(T)/c^2}
\end{align}
In a non-retarded regime, $ D_{xx}^{Sup} \propto \im[r_p] \propto (T/T_c)^4$ and as a consequence
\begin{align}
    \lim_{(T/T_c)\xrightarrow{} 0 } D_{xx}^{Sup} \approx 0 
\end{align}

\begin{figure}[t]
    \centering
  \subfloat[]{{\includegraphics[width=0.45\textwidth]{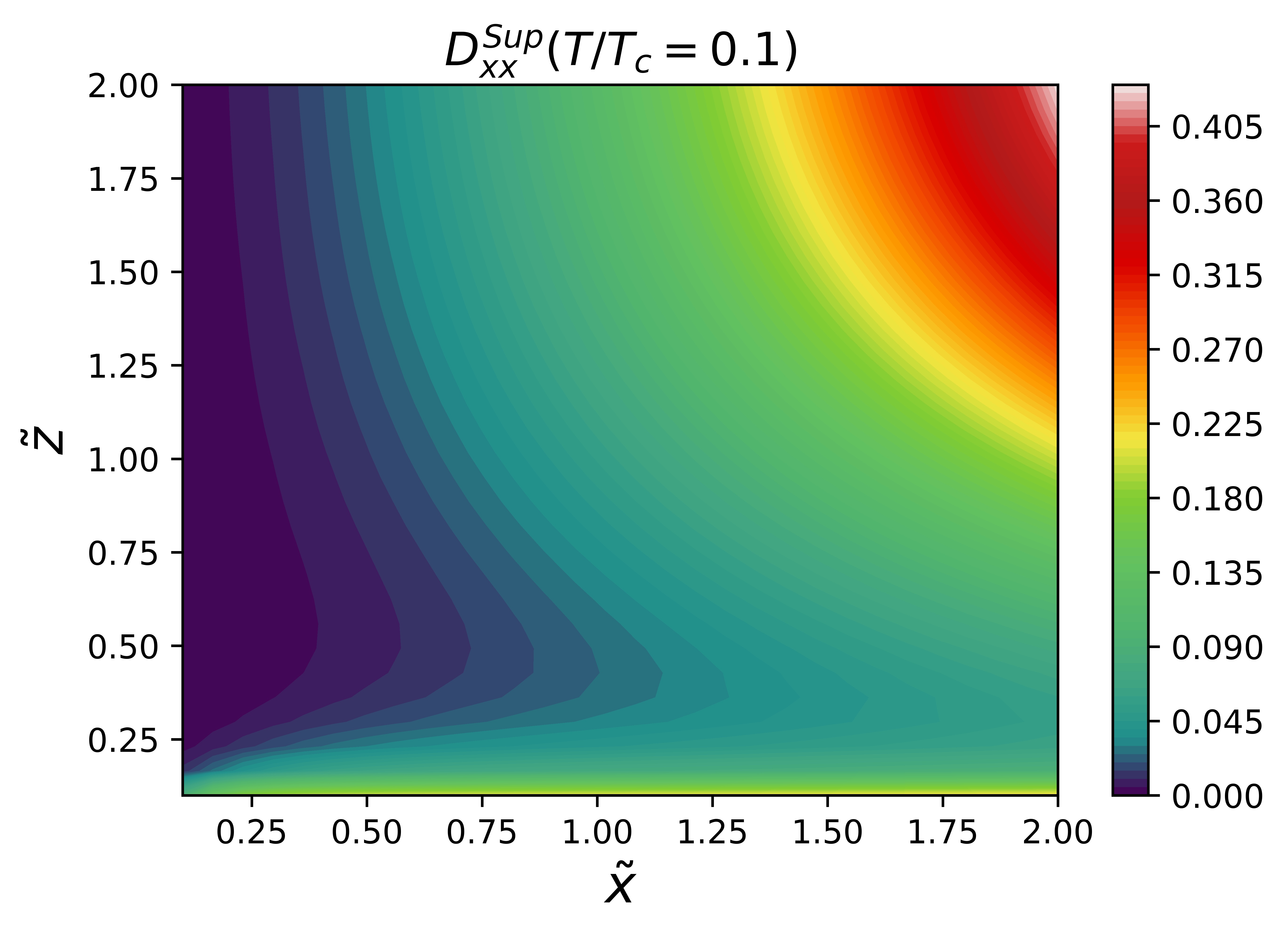} }} 
    \subfloat[]{{\includegraphics[width=0.45\textwidth]{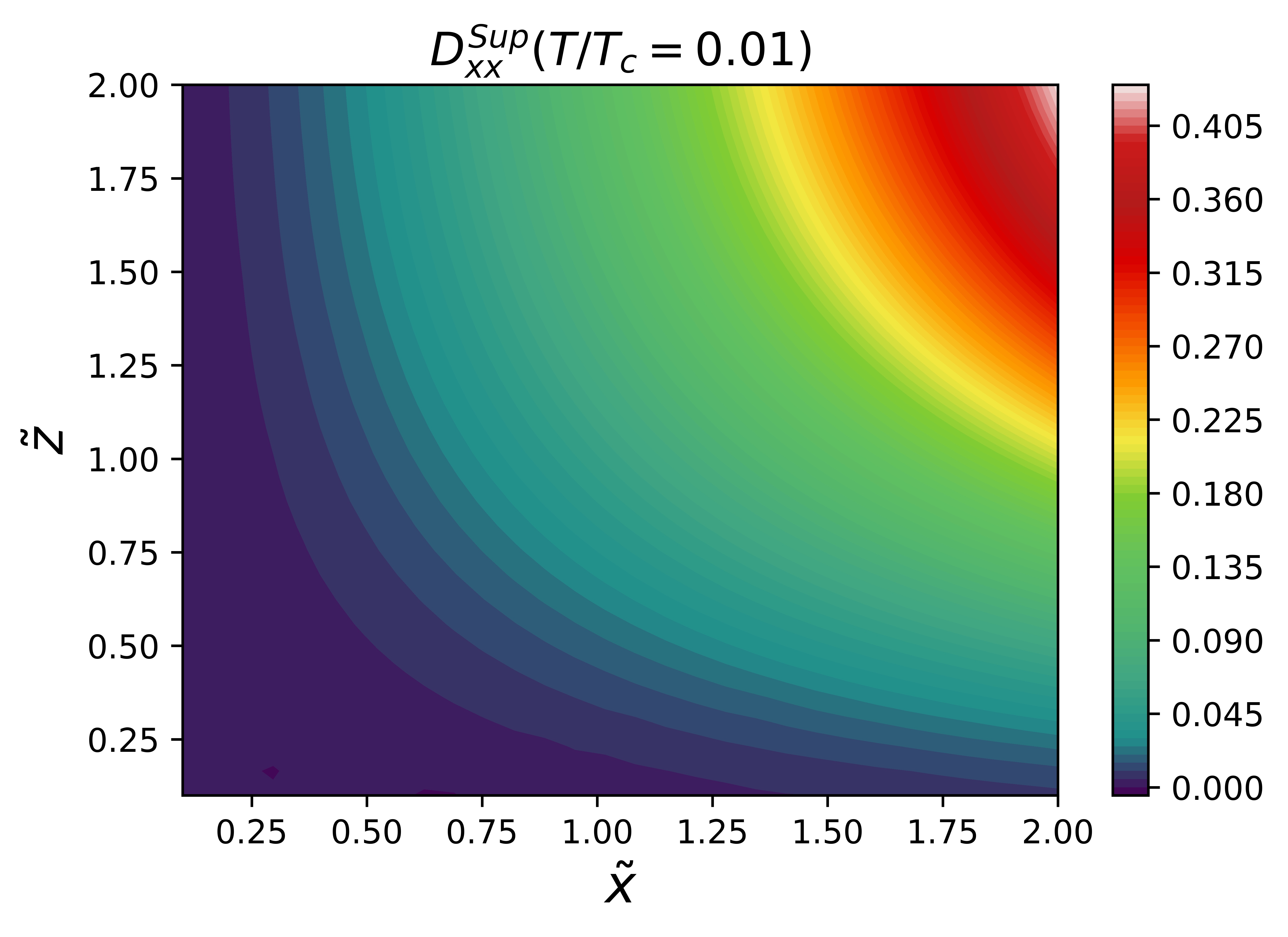} }}
     \caption{  The contour plot of the relative decay as a function of $\tilde{x}$, and $\tilde{z}$ near a Niobium superconducting surface at temperature (a) $T/T_c = 0.1 = 0.83$ K and (b) $T/T_c = 0.01 \approx 83$ mK. }
    \label{fig:Supp2}
\end{figure}
The contour plot of the relative decay is shown in Fig.~\ref{fig:Supp2} considering the Niobium superconducting surface at temperature (a) $T = 0.1~T_c$, and (b) $T = 0.01~T_c$. It displays that the relative decay near a superconducting surface behaves the same as near the perfect conductor in a low-temperature regime ($T \ll T_c$), while it tends to behave like a system near the metal surface as the temperature increases to $T \approx T_c$.


\section*{References}
\bibliography{CP}

\providecommand{\noopsort}[1]{}\providecommand{\singleletter}[1]{#1}%
\begin{thebibliography}{69}%
\makeatletter
\providecommand \@ifxundefined [1]{%
 \@ifx{#1\undefined}
}%
\providecommand \@ifnum [1]{%
 \ifnum #1\expandafter \@firstoftwo
 \else \expandafter \@secondoftwo
 \fi
}%
\providecommand \@ifx [1]{%
 \ifx #1\expandafter \@firstoftwo
 \else \expandafter \@secondoftwo
 \fi
}%
\providecommand \natexlab [1]{#1}%
\providecommand \enquote  [1]{``#1''}%
\providecommand \bibnamefont  [1]{#1}%
\providecommand \bibfnamefont [1]{#1}%
\providecommand \citenamefont [1]{#1}%
\providecommand \href@noop [0]{\@secondoftwo}%
\providecommand \href [0]{\begingroup \@sanitize@url \@href}%
\providecommand \@href[1]{\@@startlink{#1}\@@href}%
\providecommand \@@href[1]{\endgroup#1\@@endlink}%
\providecommand \@sanitize@url [0]{\catcode `\\12\catcode `\$12\catcode
  `\&12\catcode `\#12\catcode `\^12\catcode `\_12\catcode `\%12\relax}%
\providecommand \@@startlink[1]{}%
\providecommand \@@endlink[0]{}%
\providecommand \url  [0]{\begingroup\@sanitize@url \@url }%
\providecommand \@url [1]{\endgroup\@href {#1}{\urlprefix }}%
\providecommand \urlprefix  [0]{URL }%
\providecommand \Eprint [0]{\href }%
\providecommand \doibase [0]{https://doi.org/}%
\providecommand \selectlanguage [0]{\@gobble}%
\providecommand \bibinfo  [0]{\@secondoftwo}%
\providecommand \bibfield  [0]{\@secondoftwo}%
\providecommand \translation [1]{[#1]}%
\providecommand \BibitemOpen [0]{}%
\providecommand \bibitemStop [0]{}%
\providecommand \bibitemNoStop [0]{.\EOS\space}%
\providecommand \EOS [0]{\spacefactor3000\relax}%
\providecommand \BibitemShut  [1]{\csname bibitem#1\endcsname}%
\let\auto@bib@innerbib\@empty
\bibitem [{\citenamefont {D'Amico}\ and\ \citenamefont
  {et~al}(2019)}]{DamicoNQO}%
  \BibitemOpen
  \bibfield  {author} {\bibinfo {author} {\bibfnamefont {I.}~\bibnamefont
  {D'Amico}}\ and\ \bibinfo {author} {\bibnamefont {et~al}},\ }\bibfield
  {title} {\bibinfo {title} {Nanoscale quantum optics},\ }\href
  {https://doi.org/10.1393/ncr/i2019-10158-0} {\bibfield  {journal} {\bibinfo
  {journal} {La Rivista del Nuovo Cimento}\ }\textbf {\bibinfo {volume} {42}},\
  \bibinfo {pages} {153} (\bibinfo {year} {2019})}\BibitemShut {NoStop}%
\bibitem [{\citenamefont {Chang}\ \emph {et~al.}(2018)\citenamefont {Chang},
  \citenamefont {Douglas}, \citenamefont {Gonz\'alez-Tudela}, \citenamefont
  {Hung},\ and\ \citenamefont {Kimble}}]{ChangRMP2018}%
  \BibitemOpen
  \bibfield  {author} {\bibinfo {author} {\bibfnamefont {D.~E.}\ \bibnamefont
  {Chang}}, \bibinfo {author} {\bibfnamefont {J.~S.}\ \bibnamefont {Douglas}},
  \bibinfo {author} {\bibfnamefont {A.}~\bibnamefont {Gonz\'alez-Tudela}},
  \bibinfo {author} {\bibfnamefont {C.-L.}\ \bibnamefont {Hung}},\ and\
  \bibinfo {author} {\bibfnamefont {H.~J.}\ \bibnamefont {Kimble}},\ }\bibfield
   {title} {\bibinfo {title} {Colloquium: Quantum matter built from nanoscopic
  lattices of atoms and photons},\ }\href
  {https://doi.org/10.1103/RevModPhys.90.031002} {\bibfield  {journal}
  {\bibinfo  {journal} {Rev. Mod. Phys.}\ }\textbf {\bibinfo {volume} {90}},\
  \bibinfo {pages} {031002} (\bibinfo {year} {2018})}\BibitemShut {NoStop}%
\bibitem [{\citenamefont {Wang}\ \emph {et~al.}(2021)\citenamefont {Wang},
  \citenamefont {Haas},\ and\ \citenamefont {Narang}}]{Atom-field003}%
  \BibitemOpen
  \bibfield  {author} {\bibinfo {author} {\bibfnamefont {S.~D.}\ \bibnamefont
  {Wang}}, \bibinfo {author} {\bibfnamefont {M.}~\bibnamefont {Haas}},\ and\
  \bibinfo {author} {\bibfnamefont {P.}~\bibnamefont {Narang}},\ }\bibfield
  {title} {\bibinfo {title} {Quantum interfaces to the nanoscale},\ }\href
  {https://doi.org/10.1021/acsnano.1c01255} {\bibfield  {journal} {\bibinfo
  {journal} {ACS Nano}\ }\textbf {\bibinfo {volume} {15}},\ \bibinfo {pages}
  {7879} (\bibinfo {year} {2021})}\BibitemShut {NoStop}%
\bibitem [{\citenamefont {Soykal}\ and\ \citenamefont
  {Flatt\'e}(2010)}]{Atom-field004}%
  \BibitemOpen
  \bibfield  {author} {\bibinfo {author} {\bibfnamefont {O.~O.}\ \bibnamefont
  {Soykal}}\ and\ \bibinfo {author} {\bibfnamefont {M.~E.}\ \bibnamefont
  {Flatt\'e}},\ }\bibfield  {title} {\bibinfo {title} {Strong field
  interactions between a nanomagnet and a photonic cavity},\ }\href
  {https://doi.org/10.1103/PhysRevLett.104.077202} {\bibfield  {journal}
  {\bibinfo  {journal} {Phys. Rev. Lett.}\ }\textbf {\bibinfo {volume} {104}},\
  \bibinfo {pages} {077202} (\bibinfo {year} {2010})}\BibitemShut {NoStop}%
\bibitem [{\citenamefont {Gonz\'{a}lez-Tudela}\ \emph
  {et~al.}(2024)\citenamefont {Gonz\'{a}lez-Tudela}, \citenamefont {Reiserer},
  \citenamefont {Garc\'{i}a-Ripoll},\ and\ \citenamefont
  {et~al.}}]{Atom-field005}%
  \BibitemOpen
  \bibfield  {author} {\bibinfo {author} {\bibfnamefont {A.}~\bibnamefont
  {Gonz\'{a}lez-Tudela}}, \bibinfo {author} {\bibfnamefont {A.}~\bibnamefont
  {Reiserer}}, \bibinfo {author} {\bibfnamefont {J.}~\bibnamefont
  {Garc\'{i}a-Ripoll}},\ and\ \bibinfo {author} {\bibnamefont {et~al.}},\
  }\bibfield  {title} {\bibinfo {title} {Light-matter interactions in quantum
  nanophotonic devices},\ }\href {https://doi.org/10.1038/s42254-023-00681-1}
  {\bibfield  {journal} {\bibinfo  {journal} {Nat Rev Phys}\ }\textbf {\bibinfo
  {volume} {3}},\ \bibinfo {pages} {807} (\bibinfo {year} {2024})}\BibitemShut
  {NoStop}%
\bibitem [{\citenamefont {Fang}\ \emph {et~al.}(2022)\citenamefont {Fang},
  \citenamefont {Lin}, \citenamefont {xiang Li},\ and\ \citenamefont
  {Yang}}]{atom-field007}%
  \BibitemOpen
  \bibfield  {author} {\bibinfo {author} {\bibfnamefont {W.}~\bibnamefont
  {Fang}}, \bibinfo {author} {\bibfnamefont {B.}~\bibnamefont {Lin}}, \bibinfo
  {author} {\bibfnamefont {G.}~\bibnamefont {xiang Li}},\ and\ \bibinfo
  {author} {\bibfnamefont {Y.}~\bibnamefont {Yang}},\ }\bibfield  {title}
  {\bibinfo {title} {Selective mode excitations and spontaneous emission
  engineering in quantum emitter-photonic structure coupled systems},\ }\href
  {https://doi.org/10.1364/OE.455346} {\bibfield  {journal} {\bibinfo
  {journal} {Opt. Express}\ }\textbf {\bibinfo {volume} {30}},\ \bibinfo
  {pages} {21103} (\bibinfo {year} {2022})}\BibitemShut {NoStop}%
\bibitem [{\citenamefont {Chang}\ \emph
  {et~al.}(2014{\natexlab{a}})\citenamefont {Chang}, \citenamefont {Vuletic},\
  and\ \citenamefont {Lukin}}]{QNonLinear01}%
  \BibitemOpen
  \bibfield  {author} {\bibinfo {author} {\bibfnamefont {D.}~\bibnamefont
  {Chang}}, \bibinfo {author} {\bibfnamefont {V.}~\bibnamefont {Vuletic}},\
  and\ \bibinfo {author} {\bibfnamefont {M.}~\bibnamefont {Lukin}},\ }\bibfield
   {title} {\bibinfo {title} {Quantum nonlinear optics - photon by photon},\
  }\href {https://doi.org/10.1038/nphoton.2014.192} {\bibfield  {journal}
  {\bibinfo  {journal} {Nature Photon}\ }\textbf {\bibinfo {volume} {8}},\
  \bibinfo {pages} {685} (\bibinfo {year} {2014}{\natexlab{a}})}\BibitemShut
  {NoStop}%
\bibitem [{\citenamefont {Gullans}\ \emph {et~al.}(2013)\citenamefont
  {Gullans}, \citenamefont {Chang}, \citenamefont {Koppens}, \citenamefont
  {GarciadeAbajo},\ and\ \citenamefont {Lukin}}]{Nonlinear002}%
  \BibitemOpen
  \bibfield  {author} {\bibinfo {author} {\bibfnamefont {M.}~\bibnamefont
  {Gullans}}, \bibinfo {author} {\bibfnamefont {D.~E.}\ \bibnamefont {Chang}},
  \bibinfo {author} {\bibfnamefont {F.~H.~L.}\ \bibnamefont {Koppens}},
  \bibinfo {author} {\bibfnamefont {F.~J.}\ \bibnamefont {GarciadeAbajo}},\
  and\ \bibinfo {author} {\bibfnamefont {M.~D.}\ \bibnamefont {Lukin}},\
  }\bibfield  {title} {\bibinfo {title} {Single-photon nonlinear optics with
  graphene plasmons},\ }\href {https://doi.org/10.1103/PhysRevLett.111.247401}
  {\bibfield  {journal} {\bibinfo  {journal} {Phys. Rev. Lett.}\ }\textbf
  {\bibinfo {volume} {111}},\ \bibinfo {pages} {247401} (\bibinfo {year}
  {2013})}\BibitemShut {NoStop}%
\bibitem [{\citenamefont {Volz}\ \emph {et~al.}(2014)\citenamefont {Volz},
  \citenamefont {Scheucher}, \citenamefont {Junge},\ and\ \citenamefont
  {et~al.}}]{Nonlinear003}%
  \BibitemOpen
  \bibfield  {author} {\bibinfo {author} {\bibfnamefont {J.}~\bibnamefont
  {Volz}}, \bibinfo {author} {\bibfnamefont {M.}~\bibnamefont {Scheucher}},
  \bibinfo {author} {\bibfnamefont {C.}~\bibnamefont {Junge}},\ and\ \bibinfo
  {author} {\bibnamefont {et~al.}},\ }\bibfield  {title} {\bibinfo {title}
  {Nonlinear $\pi$ phase shift for single fibre-guided photons interacting with
  a single resonator-enhanced atom},\ }\href
  {https://doi.org/10.1038/nphoton.2014.253} {\bibfield  {journal} {\bibinfo
  {journal} {Nature Photon}\ }\textbf {\bibinfo {volume} {8}},\ \bibinfo
  {pages} {965} (\bibinfo {year} {2014})}\BibitemShut {NoStop}%
\bibitem [{\citenamefont {Skljarow}\ \emph {et~al.}(2022)\citenamefont
  {Skljarow}, \citenamefont {K\"ubler}, \citenamefont {Adams}, \citenamefont
  {Pfau}, \citenamefont {L\"ow},\ and\ \citenamefont {Alaeian}}]{Skljarow22}%
  \BibitemOpen
  \bibfield  {author} {\bibinfo {author} {\bibfnamefont {A.}~\bibnamefont
  {Skljarow}}, \bibinfo {author} {\bibfnamefont {H.}~\bibnamefont {K\"ubler}},
  \bibinfo {author} {\bibfnamefont {C.~S.}\ \bibnamefont {Adams}}, \bibinfo
  {author} {\bibfnamefont {T.}~\bibnamefont {Pfau}}, \bibinfo {author}
  {\bibfnamefont {R.}~\bibnamefont {L\"ow}},\ and\ \bibinfo {author}
  {\bibfnamefont {H.}~\bibnamefont {Alaeian}},\ }\bibfield  {title} {\bibinfo
  {title} {Purcell-enhanced dipolar interactions in nanostructures},\ }\href
  {https://doi.org/10.1103/PhysRevResearch.4.023073} {\bibfield  {journal}
  {\bibinfo  {journal} {Phys. Rev. Res.}\ }\textbf {\bibinfo {volume} {4}},\
  \bibinfo {pages} {023073} (\bibinfo {year} {2022})}\BibitemShut {NoStop}%
\bibitem [{\citenamefont {Goban}\ and\ \citenamefont {et~al}(2014)}]{Goban14}%
  \BibitemOpen
  \bibfield  {author} {\bibinfo {author} {\bibfnamefont {A.}~\bibnamefont
  {Goban}}\ and\ \bibinfo {author} {\bibnamefont {et~al}},\ }\bibfield  {title}
  {\bibinfo {title} {Atom--light interactions in photonic crystals},\ }\href
  {https://doi.org/10.1038/ncomms4808} {\bibfield  {journal} {\bibinfo
  {journal} {Nature Communications}\ }\textbf {\bibinfo {volume} {5}},\
  \bibinfo {pages} {3808} (\bibinfo {year} {2014})}\BibitemShut {NoStop}%
\bibitem [{\citenamefont {Thompson}\ and\ \citenamefont
  {et~al}(2013)}]{coupling011}%
  \BibitemOpen
  \bibfield  {author} {\bibinfo {author} {\bibfnamefont {J.~D.}\ \bibnamefont
  {Thompson}}\ and\ \bibinfo {author} {\bibnamefont {et~al}},\ }\bibfield
  {title} {\bibinfo {title} {Coupling a single trapped atom to a nanoscale
  optical cavity},\ }\href {https://doi.org/10.1126/science.1237125} {\bibfield
   {journal} {\bibinfo  {journal} {Science}\ }\textbf {\bibinfo {volume}
  {340}},\ \bibinfo {pages} {1202} (\bibinfo {year} {2013})}\BibitemShut
  {NoStop}%
\bibitem [{\citenamefont {Rivera}\ and\ \citenamefont
  {Kaminer}(2020)}]{interface001}%
  \BibitemOpen
  \bibfield  {author} {\bibinfo {author} {\bibfnamefont {N.}~\bibnamefont
  {Rivera}}\ and\ \bibinfo {author} {\bibfnamefont {I.}~\bibnamefont
  {Kaminer}},\ }\bibfield  {title} {\bibinfo {title} {Light--matter
  interactions with photonic quasiparticles},\ }\href
  {https://doi.org/10.1038/s42254-020-0224-2} {\bibfield  {journal} {\bibinfo
  {journal} {Nature Reviews Physics}\ }\textbf {\bibinfo {volume} {2}},\
  \bibinfo {pages} {538} (\bibinfo {year} {2020})}\BibitemShut {NoStop}%
\bibitem [{\citenamefont {Kockum}\ \emph {et~al.}(2019)\citenamefont {Kockum},
  \citenamefont {Miranowicz}, \citenamefont {Liberato},\ and\ \citenamefont
  {et~al.}}]{interface002}%
  \BibitemOpen
  \bibfield  {author} {\bibinfo {author} {\bibfnamefont {A.~F.}\ \bibnamefont
  {Kockum}}, \bibinfo {author} {\bibfnamefont {A.}~\bibnamefont {Miranowicz}},
  \bibinfo {author} {\bibfnamefont {S.~D.}\ \bibnamefont {Liberato}},\ and\
  \bibinfo {author} {\bibnamefont {et~al.}},\ }\bibfield  {title} {\bibinfo
  {title} {Ultrastrong coupling between light and matter},\ }\href
  {https://doi.org/10.1038/s42254-018-0006-2} {\bibfield  {journal} {\bibinfo
  {journal} {Nature Reviews Physics}\ }\textbf {\bibinfo {volume} {1}},\
  \bibinfo {pages} {295} (\bibinfo {year} {2019})}\BibitemShut {NoStop}%
\bibitem [{\citenamefont {Flick}\ \emph {et~al.}(2018)\citenamefont {Flick},
  \citenamefont {Rivera},\ and\ \citenamefont {Narang}}]{coupling22}%
  \BibitemOpen
  \bibfield  {author} {\bibinfo {author} {\bibfnamefont {J.}~\bibnamefont
  {Flick}}, \bibinfo {author} {\bibfnamefont {N.}~\bibnamefont {Rivera}},\ and\
  \bibinfo {author} {\bibfnamefont {P.}~\bibnamefont {Narang}},\ }\bibfield
  {title} {\bibinfo {title} {Strong light-matter coupling in quantum chemistry
  and quantum photonics},\ }\href {https://doi.org/10.1515/nanoph-2018-0067}
  {\bibfield  {journal} {\bibinfo  {journal} {Nanophotonics}\ }\textbf
  {\bibinfo {volume} {7}},\ \bibinfo {pages} {1479} (\bibinfo {year}
  {2018})}\BibitemShut {NoStop}%
\bibitem [{\citenamefont {Forn-D{\'\i}az}\ \emph {et~al.}(2019)\citenamefont
  {Forn-D{\'\i}az}, \citenamefont {Lamata}, \citenamefont {Rico}, \citenamefont
  {Kono},\ and\ \citenamefont {Solano}}]{atom-field006}%
  \BibitemOpen
  \bibfield  {author} {\bibinfo {author} {\bibfnamefont {P.}~\bibnamefont
  {Forn-D{\'\i}az}}, \bibinfo {author} {\bibfnamefont {L.}~\bibnamefont
  {Lamata}}, \bibinfo {author} {\bibfnamefont {E.}~\bibnamefont {Rico}},
  \bibinfo {author} {\bibfnamefont {J.}~\bibnamefont {Kono}},\ and\ \bibinfo
  {author} {\bibfnamefont {E.}~\bibnamefont {Solano}},\ }\bibfield  {title}
  {\bibinfo {title} {Ultrastrong coupling regimes of light-matter
  interaction},\ }\href@noop {} {\bibfield  {journal} {\bibinfo  {journal}
  {Reviews of Modern Physics}\ }\textbf {\bibinfo {volume} {91}},\ \bibinfo
  {pages} {025005} (\bibinfo {year} {2019})}\BibitemShut {NoStop}%
\bibitem [{\citenamefont {et~al}(2018{\natexlab{a}})}]{QInf001}%
  \BibitemOpen
  \bibfield  {author} {\bibinfo {author} {\bibfnamefont {F.~F.}\ \bibnamefont
  {et~al}},\ }\bibfield  {title} {\bibinfo {title} {Photonic quantum
  information processing: a review},\ }\href
  {https://doi.org/10.1088/1361-6633/aad5b2} {\bibfield  {journal} {\bibinfo
  {journal} {Reports on Progress in Physics}\ }\textbf {\bibinfo {volume}
  {82}},\ \bibinfo {pages} {016001} (\bibinfo {year}
  {2018}{\natexlab{a}})}\BibitemShut {NoStop}%
\bibitem [{\citenamefont {Laucht}\ and\ \citenamefont {et~al}(2021)}]{QInf002}%
  \BibitemOpen
  \bibfield  {author} {\bibinfo {author} {\bibfnamefont {A.}~\bibnamefont
  {Laucht}}\ and\ \bibinfo {author} {\bibnamefont {et~al}},\ }\bibfield
  {title} {\bibinfo {title} {Roadmap on quantum nanotechnologies},\ }\href
  {https://doi.org/10.1088/1361-6528/abb333} {\bibfield  {journal} {\bibinfo
  {journal} {Nanotechnology}\ }\textbf {\bibinfo {volume} {32}},\ \bibinfo
  {pages} {162003} (\bibinfo {year} {2021})}\BibitemShut {NoStop}%
\bibitem [{\citenamefont {Koo}\ \emph {et~al.}(2024)\citenamefont {Koo},
  \citenamefont {Moon}, \citenamefont {Kang},\ and\ \citenamefont
  {et~al.}}]{Ultra001}%
  \BibitemOpen
  \bibfield  {author} {\bibinfo {author} {\bibfnamefont {Y.}~\bibnamefont
  {Koo}}, \bibinfo {author} {\bibfnamefont {T.}~\bibnamefont {Moon}}, \bibinfo
  {author} {\bibfnamefont {M.}~\bibnamefont {Kang}},\ and\ \bibinfo {author}
  {\bibnamefont {et~al.}},\ }\bibfield  {title} {\bibinfo {title} {Dynamical
  control of nanoscale light-matter interactions in low-dimensional quantum
  materials},\ }\href {https://doi.org/10.1038/s41377-024-01380-x} {\bibfield
  {journal} {\bibinfo  {journal} {Light Sci Appl}\ }\textbf {\bibinfo {volume}
  {13}},\ \bibinfo {pages} {30} (\bibinfo {year} {2024})}\BibitemShut {NoStop}%
\bibitem [{\citenamefont {et~al}(2018{\natexlab{b}})}]{Ultra002}%
  \BibitemOpen
  \bibfield  {author} {\bibinfo {author} {\bibfnamefont {G.~B.}\ \bibnamefont
  {et~al}},\ }\bibfield  {title} {\bibinfo {title} {Ultrafast measurements of
  the dynamics of single nanostructures: a review},\ }\href
  {https://doi.org/10.1088/1361-6633/aaea4b} {\bibfield  {journal} {\bibinfo
  {journal} {Rep. Prog. Phys.}\ }\textbf {\bibinfo {volume} {82}},\ \bibinfo
  {pages} {016401} (\bibinfo {year} {2018}{\natexlab{b}})}\BibitemShut
  {NoStop}%
\bibitem [{\citenamefont {S\'anchez-C\'anovas}\ and\ \citenamefont
  {Donaire}(2022)}]{dipole-dipole001}%
  \BibitemOpen
  \bibfield  {author} {\bibinfo {author} {\bibfnamefont {J.}~\bibnamefont
  {S\'anchez-C\'anovas}}\ and\ \bibinfo {author} {\bibfnamefont
  {M.}~\bibnamefont {Donaire}},\ }\bibfield  {title} {\bibinfo {title}
  {Nonconservative dipole forces on an excited two-atom system},\ }\href
  {https://doi.org/10.1103/PhysRevA.106.032805} {\bibfield  {journal} {\bibinfo
   {journal} {Phys. Rev. A}\ }\textbf {\bibinfo {volume} {106}},\ \bibinfo
  {pages} {032805} (\bibinfo {year} {2022})}\BibitemShut {NoStop}%
\bibitem [{\citenamefont {Casimir}\ and\ \citenamefont
  {Polder}(1948)}]{CasimirPolder01}%
  \BibitemOpen
  \bibfield  {author} {\bibinfo {author} {\bibfnamefont {H.~B.~G.}\
  \bibnamefont {Casimir}}\ and\ \bibinfo {author} {\bibfnamefont
  {D.}~\bibnamefont {Polder}},\ }\bibfield  {title} {\bibinfo {title} {The
  influence of retardation on the london-van der waals forces},\ }\href
  {https://doi.org/10.1103/PhysRev.73.360} {\bibfield  {journal} {\bibinfo
  {journal} {Phys. Rev.}\ }\textbf {\bibinfo {volume} {73}},\ \bibinfo {pages}
  {360} (\bibinfo {year} {1948})}\BibitemShut {NoStop}%
\bibitem [{\citenamefont {Buhmann}(2012)}]{BookDispersion01}%
  \BibitemOpen
  \bibfield  {author} {\bibinfo {author} {\bibfnamefont {S.~Y.}\ \bibnamefont
  {Buhmann}},\ }\href@noop {} {\emph {\bibinfo {title} {Dispersion Forces I}}}\
  (\bibinfo  {publisher} {Springer-Verlag, Berlin},\ \bibinfo {year}
  {2012})\BibitemShut {NoStop}%
\bibitem [{\citenamefont {H\"ummer}\ \emph {et~al.}(2019)\citenamefont
  {H\"ummer}, \citenamefont {Schneeweiss}, \citenamefont {Rauschenbeutel},\
  and\ \citenamefont {Romero-Isart}}]{CPTrap01}%
  \BibitemOpen
  \bibfield  {author} {\bibinfo {author} {\bibfnamefont {D.}~\bibnamefont
  {H\"ummer}}, \bibinfo {author} {\bibfnamefont {P.}~\bibnamefont
  {Schneeweiss}}, \bibinfo {author} {\bibfnamefont {A.}~\bibnamefont
  {Rauschenbeutel}},\ and\ \bibinfo {author} {\bibfnamefont {O.}~\bibnamefont
  {Romero-Isart}},\ }\bibfield  {title} {\bibinfo {title} {Heating in
  nanophotonic traps for cold atoms},\ }\href
  {https://doi.org/10.1103/PhysRevX.9.041034} {\bibfield  {journal} {\bibinfo
  {journal} {Phys. Rev. X}\ }\textbf {\bibinfo {volume} {9}},\ \bibinfo {pages}
  {041034} (\bibinfo {year} {2019})}\BibitemShut {NoStop}%
\bibitem [{\citenamefont {Henkel}\ and\ \citenamefont
  {Wilkens}(1999)}]{CPTrap02}%
  \BibitemOpen
  \bibfield  {author} {\bibinfo {author} {\bibfnamefont {C.}~\bibnamefont
  {Henkel}}\ and\ \bibinfo {author} {\bibfnamefont {M.}~\bibnamefont
  {Wilkens}},\ }\bibfield  {title} {\bibinfo {title} {Heating of trapped atoms
  near thermal surfaces},\ }\href {https://doi.org/10.1209/epl/i1999-00404-8}
  {\bibfield  {journal} {\bibinfo  {journal} {Europhysics Letters}\ }\textbf
  {\bibinfo {volume} {47}},\ \bibinfo {pages} {414} (\bibinfo {year}
  {1999})}\BibitemShut {NoStop}%
\bibitem [{\citenamefont {Chang}\ \emph
  {et~al.}(2014{\natexlab{b}})\citenamefont {Chang}, \citenamefont {Sinha},
  \citenamefont {Taylor},\ and\ \citenamefont {Kimble}}]{Trap-Kanu02}%
  \BibitemOpen
  \bibfield  {author} {\bibinfo {author} {\bibfnamefont {D.~E.}\ \bibnamefont
  {Chang}}, \bibinfo {author} {\bibfnamefont {K.}~\bibnamefont {Sinha}},
  \bibinfo {author} {\bibfnamefont {J.~M.}\ \bibnamefont {Taylor}},\ and\
  \bibinfo {author} {\bibfnamefont {H.~J.}\ \bibnamefont {Kimble}},\ }\bibfield
   {title} {\bibinfo {title} {Trapping atoms using nanoscale quantum vacuum
  forces},\ }\href {https://doi.org/10.1038/ncomms5343} {\bibfield  {journal}
  {\bibinfo  {journal} {Nat Commun}\ }\textbf {\bibinfo {volume} {5}},\
  \bibinfo {pages} {4343} (\bibinfo {year} {2014}{\natexlab{b}})}\BibitemShut
  {NoStop}%
\bibitem [{\citenamefont {MacFarlane}\ \emph {et~al.}(2003)\citenamefont
  {MacFarlane}, \citenamefont {Dowling},\ and\ \citenamefont
  {Milburn}}]{QTech001}%
  \BibitemOpen
  \bibfield  {author} {\bibinfo {author} {\bibfnamefont {A.~G.~J.}\
  \bibnamefont {MacFarlane}}, \bibinfo {author} {\bibfnamefont {J.~P.}\
  \bibnamefont {Dowling}},\ and\ \bibinfo {author} {\bibfnamefont {G.~J.}\
  \bibnamefont {Milburn}},\ }\bibfield  {title} {\bibinfo {title} {Quantum
  technology: the second quantum revolution},\ }\href
  {https://doi.org/10.1098/rsta.2003.1227} {\bibfield  {journal} {\bibinfo
  {journal} {Philosophical Transactions of the Royal Society of London. Series
  A: Mathematical, Physical and Engineering Sciences}\ }\textbf {\bibinfo
  {volume} {361}},\ \bibinfo {pages} {1655} (\bibinfo {year}
  {2003})}\BibitemShut {NoStop}%
\bibitem [{\citenamefont {Deutsch}(2020)}]{QTech002}%
  \BibitemOpen
  \bibfield  {author} {\bibinfo {author} {\bibfnamefont {I.~H.}\ \bibnamefont
  {Deutsch}},\ }\bibfield  {title} {\bibinfo {title} {Harnessing the power of
  the second quantum revolution},\ }\href
  {https://doi.org/10.1103/PRXQuantum.1.020101} {\bibfield  {journal} {\bibinfo
   {journal} {PRX Quantum}\ }\textbf {\bibinfo {volume} {1}},\ \bibinfo {pages}
  {020101} (\bibinfo {year} {2020})}\BibitemShut {NoStop}%
\bibitem [{\citenamefont {Evans}\ and\ \citenamefont
  {et~al}(2018)}]{Control001}%
  \BibitemOpen
  \bibfield  {author} {\bibinfo {author} {\bibfnamefont {R.~E.}\ \bibnamefont
  {Evans}}\ and\ \bibinfo {author} {\bibnamefont {et~al}},\ }\bibfield  {title}
  {\bibinfo {title} {Photon-mediated interactions between quantum emitters in a
  diamond nanocavity},\ }\href {https://doi.org/10.1126/science.aau4691}
  {\bibfield  {journal} {\bibinfo  {journal} {Science}\ }\textbf {\bibinfo
  {volume} {362}},\ \bibinfo {pages} {662} (\bibinfo {year}
  {2018})}\BibitemShut {NoStop}%
\bibitem [{\citenamefont {Lukin}\ and\ \citenamefont
  {Hemmer}(2000)}]{control002}%
  \BibitemOpen
  \bibfield  {author} {\bibinfo {author} {\bibfnamefont {M.~D.}\ \bibnamefont
  {Lukin}}\ and\ \bibinfo {author} {\bibfnamefont {P.~R.}\ \bibnamefont
  {Hemmer}},\ }\bibfield  {title} {\bibinfo {title} {Quantum entanglement via
  optical control of atom-atom interactions},\ }\href
  {https://doi.org/10.1103/PhysRevLett.84.2818} {\bibfield  {journal} {\bibinfo
   {journal} {Phys. Rev. Lett.}\ }\textbf {\bibinfo {volume} {84}},\ \bibinfo
  {pages} {2818} (\bibinfo {year} {2000})}\BibitemShut {NoStop}%
\bibitem [{\citenamefont {Neuman}\ \emph {et~al.}(2020)\citenamefont {Neuman},
  \citenamefont {Trusheim},\ and\ \citenamefont {Narang}}]{control003}%
  \BibitemOpen
  \bibfield  {author} {\bibinfo {author} {\bibfnamefont {T.}~\bibnamefont
  {Neuman}}, \bibinfo {author} {\bibfnamefont {M.}~\bibnamefont {Trusheim}},\
  and\ \bibinfo {author} {\bibfnamefont {P.}~\bibnamefont {Narang}},\
  }\bibfield  {title} {\bibinfo {title} {Selective acoustic control of
  photon-mediated qubit-qubit interactions},\ }\href
  {https://doi.org/10.1103/PhysRevA.101.052342} {\bibfield  {journal} {\bibinfo
   {journal} {Phys. Rev. A}\ }\textbf {\bibinfo {volume} {101}},\ \bibinfo
  {pages} {052342} (\bibinfo {year} {2020})}\BibitemShut {NoStop}%
\bibitem [{\citenamefont {Laucht}\ and\ \citenamefont
  {et~al}(2009)}]{control004}%
  \BibitemOpen
  \bibfield  {author} {\bibinfo {author} {\bibfnamefont {A.}~\bibnamefont
  {Laucht}}\ and\ \bibinfo {author} {\bibnamefont {et~al}},\ }\bibfield
  {title} {\bibinfo {title} {Electrical control of spontaneous emission and
  strong coupling for a single quantum dot},\ }\href
  {https://doi.org/https://iopscience.iop.org/article/10.1088/1367-2630/11/2/023034}
  {\bibfield  {journal} {\bibinfo  {journal} {New Journal of Physics}\ }\textbf
  {\bibinfo {volume} {11}},\ \bibinfo {pages} {023034} (\bibinfo {year}
  {2009})}\BibitemShut {NoStop}%
\bibitem [{\citenamefont {Braun}(2002)}]{BathEnt01}%
  \BibitemOpen
  \bibfield  {author} {\bibinfo {author} {\bibfnamefont {D.}~\bibnamefont
  {Braun}},\ }\bibfield  {title} {\bibinfo {title} {Creation of entanglement by
  interaction with a common heat bath},\ }\href
  {https://doi.org/10.1103/PhysRevLett.89.277901} {\bibfield  {journal}
  {\bibinfo  {journal} {Phys. Rev. Lett.}\ }\textbf {\bibinfo {volume} {89}},\
  \bibinfo {pages} {277901} (\bibinfo {year} {2002})}\BibitemShut {NoStop}%
\bibitem [{\citenamefont {Benatti}\ \emph {et~al.}(2003)\citenamefont
  {Benatti}, \citenamefont {Floreanini},\ and\ \citenamefont
  {Piani}}]{BathEnt02}%
  \BibitemOpen
  \bibfield  {author} {\bibinfo {author} {\bibfnamefont {F.}~\bibnamefont
  {Benatti}}, \bibinfo {author} {\bibfnamefont {R.}~\bibnamefont
  {Floreanini}},\ and\ \bibinfo {author} {\bibfnamefont {M.}~\bibnamefont
  {Piani}},\ }\bibfield  {title} {\bibinfo {title} {Environment induced
  entanglement in markovian dissipative dynamics},\ }\href
  {https://doi.org/10.1103/PhysRevLett.91.070402} {\bibfield  {journal}
  {\bibinfo  {journal} {Phys. Rev. Lett.}\ }\textbf {\bibinfo {volume} {91}},\
  \bibinfo {pages} {070402} (\bibinfo {year} {2003})}\BibitemShut {NoStop}%
\bibitem [{\citenamefont {Cattaneo}\ \emph {et~al.}(2021)\citenamefont
  {Cattaneo}, \citenamefont {Giorgi}, \citenamefont {Maniscalco}, \citenamefont
  {Paraoanu},\ and\ \citenamefont {Zambrini}}]{BathEnt03}%
  \BibitemOpen
  \bibfield  {author} {\bibinfo {author} {\bibfnamefont {M.}~\bibnamefont
  {Cattaneo}}, \bibinfo {author} {\bibfnamefont {G.~L.}\ \bibnamefont
  {Giorgi}}, \bibinfo {author} {\bibfnamefont {S.}~\bibnamefont {Maniscalco}},
  \bibinfo {author} {\bibfnamefont {G.~S.}\ \bibnamefont {Paraoanu}},\ and\
  \bibinfo {author} {\bibfnamefont {R.}~\bibnamefont {Zambrini}},\ }\bibfield
  {title} {\bibinfo {title} {Bath-induced collective phenomena on
  superconducting qubits: Synchronization, subradiance, and entanglement
  generation},\ }\href {https://doi.org/https://doi.org/10.1002/andp.202100038}
  {\bibfield  {journal} {\bibinfo  {journal} {Annalen der Physik}\ }\textbf
  {\bibinfo {volume} {533}},\ \bibinfo {pages} {2100038} (\bibinfo {year}
  {2021})}\BibitemShut {NoStop}%
\bibitem [{\citenamefont {Dung}\ \emph {et~al.}(2002)\citenamefont {Dung},
  \citenamefont {Scheel}, \citenamefont {Welsch},\ and\ \citenamefont
  {Kn\"{o}ll}}]{ScheelMain}%
  \BibitemOpen
  \bibfield  {author} {\bibinfo {author} {\bibfnamefont {H.~T.}\ \bibnamefont
  {Dung}}, \bibinfo {author} {\bibfnamefont {S.}~\bibnamefont {Scheel}},
  \bibinfo {author} {\bibfnamefont {D.-G.}\ \bibnamefont {Welsch}},\ and\
  \bibinfo {author} {\bibfnamefont {L.}~\bibnamefont {Kn\"{o}ll}},\ }\bibfield
  {title} {\bibinfo {title} {Atomic entanglement near a realistic
  microsphere},\ }\href {https://doi.org/10.1088/1464-4266/4/3/371} {\bibfield
  {journal} {\bibinfo  {journal} {Journal of Optics B: Quantum and
  Semiclassical Optics}\ }\textbf {\bibinfo {volume} {4}},\ \bibinfo {pages}
  {S169} (\bibinfo {year} {2002})}\BibitemShut {NoStop}%
\bibitem [{\citenamefont {Amooghorban}\ and\ \citenamefont
  {Aleebrahim}(2017)}]{Cas_Ent01}%
  \BibitemOpen
  \bibfield  {author} {\bibinfo {author} {\bibfnamefont {E.}~\bibnamefont
  {Amooghorban}}\ and\ \bibinfo {author} {\bibfnamefont {E.}~\bibnamefont
  {Aleebrahim}},\ }\bibfield  {title} {\bibinfo {title} {Entanglement dynamics
  of two two-level atoms in the vicinity of an invisibility cloak},\ }\href
  {https://doi.org/10.1103/PhysRevA.96.012339} {\bibfield  {journal} {\bibinfo
  {journal} {Phys. Rev. A}\ }\textbf {\bibinfo {volume} {96}},\ \bibinfo
  {pages} {012339} (\bibinfo {year} {2017})}\BibitemShut {NoStop}%
\bibitem [{\citenamefont {Brownnutt}\ \emph {et~al.}(2015)\citenamefont
  {Brownnutt}, \citenamefont {Kumph}, \citenamefont {Rabl},\ and\ \citenamefont
  {Blatt}}]{Noise001}%
  \BibitemOpen
  \bibfield  {author} {\bibinfo {author} {\bibfnamefont {M.}~\bibnamefont
  {Brownnutt}}, \bibinfo {author} {\bibfnamefont {M.}~\bibnamefont {Kumph}},
  \bibinfo {author} {\bibfnamefont {P.}~\bibnamefont {Rabl}},\ and\ \bibinfo
  {author} {\bibfnamefont {R.}~\bibnamefont {Blatt}},\ }\bibfield  {title}
  {\bibinfo {title} {Ion-trap measurements of electric-field noise near
  surfaces},\ }\href {https://doi.org/10.1103/RevModPhys.87.1419} {\bibfield
  {journal} {\bibinfo  {journal} {Rev. Mod. Phys.}\ }\textbf {\bibinfo {volume}
  {87}},\ \bibinfo {pages} {1419} (\bibinfo {year} {2015})}\BibitemShut
  {NoStop}%
\bibitem [{\citenamefont {Wang}\ \emph {et~al.}(2015)\citenamefont {Wang},
  \citenamefont {Axline}, \citenamefont {Gao}, \citenamefont {Brecht},
  \citenamefont {Chu}, \citenamefont {Frunzio}, \citenamefont {Devoret},\ and\
  \citenamefont {Schoelkopf}}]{Noise002}%
  \BibitemOpen
  \bibfield  {author} {\bibinfo {author} {\bibfnamefont {C.}~\bibnamefont
  {Wang}}, \bibinfo {author} {\bibfnamefont {C.}~\bibnamefont {Axline}},
  \bibinfo {author} {\bibfnamefont {Y.~Y.}\ \bibnamefont {Gao}}, \bibinfo
  {author} {\bibfnamefont {T.}~\bibnamefont {Brecht}}, \bibinfo {author}
  {\bibfnamefont {Y.}~\bibnamefont {Chu}}, \bibinfo {author} {\bibfnamefont
  {L.}~\bibnamefont {Frunzio}}, \bibinfo {author} {\bibfnamefont {M.~H.}\
  \bibnamefont {Devoret}},\ and\ \bibinfo {author} {\bibfnamefont {R.~J.}\
  \bibnamefont {Schoelkopf}},\ }\bibfield  {title} {\bibinfo {title} {{Surface
  participation and dielectric loss in superconducting qubits}},\ }\href
  {https://doi.org/10.1063/1.4934486} {\bibfield  {journal} {\bibinfo
  {journal} {Applied Physics Letters}\ }\textbf {\bibinfo {volume} {107}},\
  \bibinfo {pages} {162601} (\bibinfo {year} {2015})}\BibitemShut {NoStop}%
\bibitem [{\citenamefont {Kim}\ \emph {et~al.}(2015)\citenamefont {Kim},
  \citenamefont {Mamin}, \citenamefont {Sherwood}, \citenamefont {Ohno},
  \citenamefont {Awschalom},\ and\ \citenamefont {Rugar}}]{Noise003}%
  \BibitemOpen
  \bibfield  {author} {\bibinfo {author} {\bibfnamefont {M.}~\bibnamefont
  {Kim}}, \bibinfo {author} {\bibfnamefont {H.~J.}\ \bibnamefont {Mamin}},
  \bibinfo {author} {\bibfnamefont {M.~H.}\ \bibnamefont {Sherwood}}, \bibinfo
  {author} {\bibfnamefont {K.}~\bibnamefont {Ohno}}, \bibinfo {author}
  {\bibfnamefont {D.~D.}\ \bibnamefont {Awschalom}},\ and\ \bibinfo {author}
  {\bibfnamefont {D.}~\bibnamefont {Rugar}},\ }\bibfield  {title} {\bibinfo
  {title} {Decoherence of near-surface nitrogen-vacancy centers due to electric
  field noise},\ }\href {https://doi.org/10.1103/PhysRevLett.115.087602}
  {\bibfield  {journal} {\bibinfo  {journal} {Phys. Rev. Lett.}\ }\textbf
  {\bibinfo {volume} {115}},\ \bibinfo {pages} {087602} (\bibinfo {year}
  {2015})}\BibitemShut {NoStop}%
\bibitem [{\citenamefont {Jamonneau}\ and\ \citenamefont
  {et~al}(2016)}]{Noise004}%
  \BibitemOpen
  \bibfield  {author} {\bibinfo {author} {\bibfnamefont {P.}~\bibnamefont
  {Jamonneau}}\ and\ \bibinfo {author} {\bibnamefont {et~al}},\ }\bibfield
  {title} {\bibinfo {title} {Competition between electric field and magnetic
  field noise in the decoherence of a single spin in diamond},\ }\href
  {https://doi.org/10.1103/PhysRevB.93.024305} {\bibfield  {journal} {\bibinfo
  {journal} {Phys. Rev. B}\ }\textbf {\bibinfo {volume} {93}},\ \bibinfo
  {pages} {024305} (\bibinfo {year} {2016})}\BibitemShut {NoStop}%
\bibitem [{\citenamefont {Yang}\ \emph {et~al.}(2011)\citenamefont {Yang},
  \citenamefont {Callegari}, \citenamefont {Feng},\ and\ \citenamefont
  {Roukes}}]{Noise005}%
  \BibitemOpen
  \bibfield  {author} {\bibinfo {author} {\bibfnamefont {Y.}~\bibnamefont
  {Yang}}, \bibinfo {author} {\bibfnamefont {C.}~\bibnamefont {Callegari}},
  \bibinfo {author} {\bibfnamefont {X.}~\bibnamefont {Feng}},\ and\ \bibinfo
  {author} {\bibfnamefont {M.}~\bibnamefont {Roukes}},\ }\bibfield  {title}
  {\bibinfo {title} {Surface adsorbate fluctuations and noise in
  nanoelectromechanical systems},\ }\href {https://doi.org/10.1021/nl2003158}
  {\bibfield  {journal} {\bibinfo  {journal} {Nano letters}\ }\textbf {\bibinfo
  {volume} {11}},\ \bibinfo {pages} {1753} (\bibinfo {year}
  {2011})}\BibitemShut {NoStop}%
\bibitem [{\citenamefont {Reiche}\ \emph {et~al.}(2020)\citenamefont {Reiche},
  \citenamefont {Busch},\ and\ \citenamefont {Intravaia}}]{Noise006}%
  \BibitemOpen
  \bibfield  {author} {\bibinfo {author} {\bibfnamefont {D.}~\bibnamefont
  {Reiche}}, \bibinfo {author} {\bibfnamefont {K.}~\bibnamefont {Busch}},\ and\
  \bibinfo {author} {\bibfnamefont {F.}~\bibnamefont {Intravaia}},\ }\bibfield
  {title} {\bibinfo {title} {Nonadditive enhancement of nonequilibrium
  atom-surface interactions},\ }\href
  {https://doi.org/10.1103/PhysRevLett.124.193603} {\bibfield  {journal}
  {\bibinfo  {journal} {Phys. Rev. Lett.}\ }\textbf {\bibinfo {volume} {124}},\
  \bibinfo {pages} {193603} (\bibinfo {year} {2020})}\BibitemShut {NoStop}%
\bibitem [{\citenamefont {Jain}\ \emph {et~al.}(2024)\citenamefont {Jain},
  \citenamefont {Ruks}, \citenamefont {le~Kien},\ and\ \citenamefont
  {Busch}}]{Dipole-Dipole2024}%
  \BibitemOpen
  \bibfield  {author} {\bibinfo {author} {\bibfnamefont {K.}~\bibnamefont
  {Jain}}, \bibinfo {author} {\bibfnamefont {L.}~\bibnamefont {Ruks}}, \bibinfo
  {author} {\bibfnamefont {F.}~\bibnamefont {le~Kien}},\ and\ \bibinfo {author}
  {\bibfnamefont {T.}~\bibnamefont {Busch}},\ }\href@noop {} {\bibinfo {title}
  {Strong dipole-dipole interactions via enhanced light-matter coupling in
  composite nanofiber waveguides}},\ \bibinfo {howpublished} {e-print
  arXiv:2405.06168 [quant-ph]} (\bibinfo {year} {2024})\BibitemShut {NoStop}%
\bibitem [{\citenamefont {Donaire}\ and\ \citenamefont
  {Lambrecht}(2015)}]{Lambrecht2015}%
  \BibitemOpen
  \bibfield  {author} {\bibinfo {author} {\bibfnamefont {M.}~\bibnamefont
  {Donaire}}\ and\ \bibinfo {author} {\bibfnamefont {A.}~\bibnamefont
  {Lambrecht}},\ }\bibfield  {title} {\bibinfo {title} {Coherent effect of
  vacuum fluctuations on driven atoms},\ }\href
  {https://doi.org/10.1103/PhysRevA.92.013838} {\bibfield  {journal} {\bibinfo
  {journal} {Phys. Rev. A}\ }\textbf {\bibinfo {volume} {92}},\ \bibinfo
  {pages} {013838} (\bibinfo {year} {2015})}\BibitemShut {NoStop}%
\bibitem [{\citenamefont {Donaire}\ \emph {et~al.}(2015)\citenamefont
  {Donaire}, \citenamefont {Gorza}, \citenamefont {Maury}, \citenamefont
  {Guerout},\ and\ \citenamefont {Lambrecht}}]{Donaire_2015}%
  \BibitemOpen
  \bibfield  {author} {\bibinfo {author} {\bibfnamefont {M.}~\bibnamefont
  {Donaire}}, \bibinfo {author} {\bibfnamefont {M.}~\bibnamefont {Gorza}},
  \bibinfo {author} {\bibfnamefont {A.}~\bibnamefont {Maury}}, \bibinfo
  {author} {\bibfnamefont {R.}~\bibnamefont {Guerout}},\ and\ \bibinfo {author}
  {\bibfnamefont {A.}~\bibnamefont {Lambrecht}},\ }\bibfield  {title} {\bibinfo
  {title} {Casimir-polder-induced rabi oscillations},\ }\href
  {https://doi.org/10.1209/0295-5075/109/24003} {\bibfield  {journal} {\bibinfo
   {journal} {Europhysics Letters}\ }\textbf {\bibinfo {volume} {109}},\
  \bibinfo {pages} {24003} (\bibinfo {year} {2015})}\BibitemShut {NoStop}%
\bibitem [{\citenamefont {Scheel}\ and\ \citenamefont
  {Buhmann}(2009)}]{BathEnt04}%
  \BibitemOpen
  \bibfield  {author} {\bibinfo {author} {\bibfnamefont {S.}~\bibnamefont
  {Scheel}}\ and\ \bibinfo {author} {\bibfnamefont {S.~Y.}\ \bibnamefont
  {Buhmann}},\ }\href@noop {} {\bibinfo {title} {Macroscopic qed - concepts and
  applications}} (\bibinfo {year} {2009}),\ \Eprint
  {https://arxiv.org/abs/0902.3586} {arXiv:0902.3586 [quant-ph]} \BibitemShut
  {NoStop}%
\bibitem [{\citenamefont {Breuer}\ and\ \citenamefont
  {Petruccione}(2002)}]{BookMaster03}%
  \BibitemOpen
  \bibfield  {author} {\bibinfo {author} {\bibfnamefont {H.-P.}\ \bibnamefont
  {Breuer}}\ and\ \bibinfo {author} {\bibfnamefont {F.}~\bibnamefont
  {Petruccione}},\ }\href@noop {} {\emph {\bibinfo {title} {Theory of Open
  Quantum Systems}}}\ (\bibinfo  {publisher} {Oxford University Press, New
  York},\ \bibinfo {year} {2002})\BibitemShut {NoStop}%
\bibitem [{\citenamefont {Sinha}\ \emph {et~al.}(2018)\citenamefont {Sinha},
  \citenamefont {Venkatesh},\ and\ \citenamefont
  {Meystre}}]{Collective-Kanu01}%
  \BibitemOpen
  \bibfield  {author} {\bibinfo {author} {\bibfnamefont {K.}~\bibnamefont
  {Sinha}}, \bibinfo {author} {\bibfnamefont {B.~P.}\ \bibnamefont
  {Venkatesh}},\ and\ \bibinfo {author} {\bibfnamefont {P.}~\bibnamefont
  {Meystre}},\ }\bibfield  {title} {\bibinfo {title} {Collective effects in
  casimir-polder forces},\ }\href
  {https://doi.org/10.1103/PhysRevLett.121.183605} {\bibfield  {journal}
  {\bibinfo  {journal} {Phys. Rev. Lett}\ }\textbf {\bibinfo {volume} {121}},\
  \bibinfo {pages} {183605} (\bibinfo {year} {2018})}\BibitemShut {NoStop}%
\bibitem [{\citenamefont {Olivera}\ \emph {et~al.}(2022)\citenamefont
  {Olivera}, \citenamefont {Sinha},\ and\ \citenamefont {Solano}}]{Olivera22}%
  \BibitemOpen
  \bibfield  {author} {\bibinfo {author} {\bibfnamefont {A.}~\bibnamefont
  {Olivera}}, \bibinfo {author} {\bibfnamefont {K.}~\bibnamefont {Sinha}},\
  and\ \bibinfo {author} {\bibfnamefont {P.}~\bibnamefont {Solano}},\
  }\bibfield  {title} {\bibinfo {title} {Dipole-dipole interactions through a
  lens},\ }\href {https://doi.org/10.1103/PhysRevA.106.013703} {\bibfield
  {journal} {\bibinfo  {journal} {Phys. Rev. A}\ }\textbf {\bibinfo {volume}
  {106}},\ \bibinfo {pages} {013703} (\bibinfo {year} {2022})}\BibitemShut
  {NoStop}%
\bibitem [{\citenamefont {Sone}\ \emph {et~al.}(2024)\citenamefont {Sone},
  \citenamefont {Sinha},\ and\ \citenamefont {Deffner}}]{sone2024}%
  \BibitemOpen
  \bibfield  {author} {\bibinfo {author} {\bibfnamefont {A.}~\bibnamefont
  {Sone}}, \bibinfo {author} {\bibfnamefont {K.}~\bibnamefont {Sinha}},\ and\
  \bibinfo {author} {\bibfnamefont {S.}~\bibnamefont {Deffner}},\ }\href@noop
  {} {\bibinfo {title} {Thermodynamic perspective on quantum fluctuations}},\
  \bibinfo {howpublished} {e-print arXiv:2308.04951 [cond-mat.stat-mech]}
  (\bibinfo {year} {2024})\BibitemShut {NoStop}%
\bibitem [{\citenamefont {Dicke}(1954)}]{Dicke}%
  \BibitemOpen
  \bibfield  {author} {\bibinfo {author} {\bibfnamefont {R.~H.}\ \bibnamefont
  {Dicke}},\ }\bibfield  {title} {\bibinfo {title} {Coherence in spontaneous
  radiation processes},\ }\href {https://doi.org/10.1103/PhysRev.93.99}
  {\bibfield  {journal} {\bibinfo  {journal} {Phys. Rev.}\ }\textbf {\bibinfo
  {volume} {93}},\ \bibinfo {pages} {99} (\bibinfo {year} {1954})}\BibitemShut
  {NoStop}%
\bibitem [{\citenamefont {Hill}\ and\ \citenamefont
  {Wootters}(1997)}]{concurrence01}%
  \BibitemOpen
  \bibfield  {author} {\bibinfo {author} {\bibfnamefont {S.~A.}\ \bibnamefont
  {Hill}}\ and\ \bibinfo {author} {\bibfnamefont {W.~K.}\ \bibnamefont
  {Wootters}},\ }\bibfield  {title} {\bibinfo {title} {Entanglement of a pair
  of quantum bits},\ }\href {https://doi.org/10.1103/PhysRevLett.78.5022}
  {\bibfield  {journal} {\bibinfo  {journal} {Phys. Rev. Lett.}\ }\textbf
  {\bibinfo {volume} {78}},\ \bibinfo {pages} {5022} (\bibinfo {year}
  {1997})}\BibitemShut {NoStop}%
\bibitem [{\citenamefont {Wootters}(1998)}]{concurrence02}%
  \BibitemOpen
  \bibfield  {author} {\bibinfo {author} {\bibfnamefont {W.~K.}\ \bibnamefont
  {Wootters}},\ }\bibfield  {title} {\bibinfo {title} {Entanglement of
  formation of an arbitrary state of two qubits},\ }\href
  {https://doi.org/10.1103/PhysRevLett.80.2245} {\bibfield  {journal} {\bibinfo
   {journal} {Phys. Rev. Lett.}\ }\textbf {\bibinfo {volume} {80}},\ \bibinfo
  {pages} {2245} (\bibinfo {year} {1998})}\BibitemShut {NoStop}%
\bibitem [{\citenamefont {Horodecki}\ \emph {et~al.}(2009)\citenamefont
  {Horodecki}, \citenamefont {Horodecki}, \citenamefont {Horodecki},\ and\
  \citenamefont {Horodecki}}]{ConcurrenceRef}%
  \BibitemOpen
  \bibfield  {author} {\bibinfo {author} {\bibfnamefont {R.}~\bibnamefont
  {Horodecki}}, \bibinfo {author} {\bibfnamefont {P.}~\bibnamefont
  {Horodecki}}, \bibinfo {author} {\bibfnamefont {M.}~\bibnamefont
  {Horodecki}},\ and\ \bibinfo {author} {\bibfnamefont {K.}~\bibnamefont
  {Horodecki}},\ }\bibfield  {title} {\bibinfo {title} {Quantum entanglement},\
  }\href {https://doi.org/10.1103/RevModPhys.81.865} {\bibfield  {journal}
  {\bibinfo  {journal} {Rev. Mod. Phys.}\ }\textbf {\bibinfo {volume} {81}},\
  \bibinfo {pages} {865} (\bibinfo {year} {2009})}\BibitemShut {NoStop}%
\bibitem [{\citenamefont {Zhou}\ \emph {et~al.}(2017)\citenamefont {Zhou},
  \citenamefont {Rasmita}, \citenamefont {Li},\ and\ \citenamefont
  {et~al.}}]{NVSi01}%
  \BibitemOpen
  \bibfield  {author} {\bibinfo {author} {\bibfnamefont {Y.}~\bibnamefont
  {Zhou}}, \bibinfo {author} {\bibfnamefont {A.}~\bibnamefont {Rasmita}},
  \bibinfo {author} {\bibfnamefont {K.}~\bibnamefont {Li}},\ and\ \bibinfo
  {author} {\bibnamefont {et~al.}},\ }\bibfield  {title} {\bibinfo {title}
  {Coherent control of a strongly driven silicon vacancy optical transition in
  diamond},\ }\href {https://doi.org/10.1038/ncomms14451} {\bibfield  {journal}
  {\bibinfo  {journal} {Nature Photon}\ }\textbf {\bibinfo {volume} {8}},\
  \bibinfo {pages} {14451} (\bibinfo {year} {2017})}\BibitemShut {NoStop}%
\bibitem [{\citenamefont {Pirozhenko}\ \emph {et~al.}(2006)\citenamefont
  {Pirozhenko}, \citenamefont {Lambrecht},\ and\ \citenamefont
  {Svetovoy}}]{GoldRef}%
  \BibitemOpen
  \bibfield  {author} {\bibinfo {author} {\bibfnamefont {I.}~\bibnamefont
  {Pirozhenko}}, \bibinfo {author} {\bibfnamefont {A.}~\bibnamefont
  {Lambrecht}},\ and\ \bibinfo {author} {\bibfnamefont {V.~B.}\ \bibnamefont
  {Svetovoy}},\ }\bibfield  {title} {\bibinfo {title} {Sample dependence of the
  casimir force},\ }\href {https://doi.org/10.1088/1367-2630/8/10/238}
  {\bibfield  {journal} {\bibinfo  {journal} {New Journal of Physics}\ }\textbf
  {\bibinfo {volume} {8}},\ \bibinfo {pages} {238} (\bibinfo {year}
  {2006})}\BibitemShut {NoStop}%
\bibitem [{\citenamefont {Jackson}(1999)}]{BookJackson}%
  \BibitemOpen
  \bibfield  {author} {\bibinfo {author} {\bibfnamefont {J.~D.}\ \bibnamefont
  {Jackson}},\ }\href@noop {} {\emph {\bibinfo {title} {Classical
  electrodynamics}}}\ (\bibinfo  {publisher} {Wiley, New York},\ \bibinfo
  {year} {1999})\BibitemShut {NoStop}%
\bibitem [{\citenamefont {Skagerstam}\ \emph {et~al.}(2006)\citenamefont
  {Skagerstam}, \citenamefont {Hohenester}, \citenamefont {Eiguren},\ and\
  \citenamefont {Rekdal}}]{Superconducting001}%
  \BibitemOpen
  \bibfield  {author} {\bibinfo {author} {\bibfnamefont {B.-S.~K.}\
  \bibnamefont {Skagerstam}}, \bibinfo {author} {\bibfnamefont
  {U.}~\bibnamefont {Hohenester}}, \bibinfo {author} {\bibfnamefont
  {A.}~\bibnamefont {Eiguren}},\ and\ \bibinfo {author} {\bibfnamefont {P.~K.}\
  \bibnamefont {Rekdal}},\ }\bibfield  {title} {\bibinfo {title} {Spin
  decoherence in superconducting atom chips},\ }\href
  {https://doi.org/10.1103/PhysRevLett.97.070401} {\bibfield  {journal}
  {\bibinfo  {journal} {Phys. Rev. Lett.}\ }\textbf {\bibinfo {volume} {97}},\
  \bibinfo {pages} {070401} (\bibinfo {year} {2006})}\BibitemShut {NoStop}%
\bibitem [{\citenamefont {Rodriguez-Lopez}\ \emph {et~al.}(2015)\citenamefont
  {Rodriguez-Lopez}, \citenamefont {Emig}, \citenamefont {Noruzifar},\ and\
  \citenamefont {Zandi}}]{CurvaturePRA01}%
  \BibitemOpen
  \bibfield  {author} {\bibinfo {author} {\bibfnamefont {P.}~\bibnamefont
  {Rodriguez-Lopez}}, \bibinfo {author} {\bibfnamefont {T.}~\bibnamefont
  {Emig}}, \bibinfo {author} {\bibfnamefont {E.}~\bibnamefont {Noruzifar}},\
  and\ \bibinfo {author} {\bibfnamefont {R.}~\bibnamefont {Zandi}},\ }\bibfield
   {title} {\bibinfo {title} {Effect of curvature and confinement on the
  casimir-polder interaction},\ }\href
  {https://doi.org/10.1103/PhysRevA.91.012516} {\bibfield  {journal} {\bibinfo
  {journal} {Phys. Rev. A}\ }\textbf {\bibinfo {volume} {91}},\ \bibinfo
  {pages} {012516} (\bibinfo {year} {2015})}\BibitemShut {NoStop}%
\bibitem [{\citenamefont {Asenjo-Garcia}\ \emph {et~al.}(2017)\citenamefont
  {Asenjo-Garcia}, \citenamefont {Moreno-Cardoner}, \citenamefont {Albrecht},
  \citenamefont {Kimble},\ and\ \citenamefont {Chang}}]{NanoFiber01}%
  \BibitemOpen
  \bibfield  {author} {\bibinfo {author} {\bibfnamefont {A.}~\bibnamefont
  {Asenjo-Garcia}}, \bibinfo {author} {\bibfnamefont {M.}~\bibnamefont
  {Moreno-Cardoner}}, \bibinfo {author} {\bibfnamefont {A.}~\bibnamefont
  {Albrecht}}, \bibinfo {author} {\bibfnamefont {H.~J.}\ \bibnamefont
  {Kimble}},\ and\ \bibinfo {author} {\bibfnamefont {D.~E.}\ \bibnamefont
  {Chang}},\ }\bibfield  {title} {\bibinfo {title} {Exponential improvement in
  photon storage fidelities using subradiance and ``selective radiance'' in
  atomic arrays},\ }\href {https://doi.org/10.1103/PhysRevX.7.031024}
  {\bibfield  {journal} {\bibinfo  {journal} {Phys. Rev. X}\ }\textbf {\bibinfo
  {volume} {7}},\ \bibinfo {pages} {031024} (\bibinfo {year}
  {2017})}\BibitemShut {NoStop}%
\bibitem [{\citenamefont {Bose}\ \emph {et~al.}(2017)\citenamefont {Bose},
  \citenamefont {Mazumdar}, \citenamefont {Morley}, \citenamefont {Ulbricht},
  \citenamefont {Toro\ifmmode~\check{s}\else \v{s}\fi{}}, \citenamefont
  {Paternostro}, \citenamefont {Geraci}, \citenamefont {Barker}, \citenamefont
  {Kim},\ and\ \citenamefont {Milburn}}]{Anupam01}%
  \BibitemOpen
  \bibfield  {author} {\bibinfo {author} {\bibfnamefont {S.}~\bibnamefont
  {Bose}}, \bibinfo {author} {\bibfnamefont {A.}~\bibnamefont {Mazumdar}},
  \bibinfo {author} {\bibfnamefont {G.~W.}\ \bibnamefont {Morley}}, \bibinfo
  {author} {\bibfnamefont {H.}~\bibnamefont {Ulbricht}}, \bibinfo {author}
  {\bibfnamefont {M.}~\bibnamefont {Toro\ifmmode~\check{s}\else \v{s}\fi{}}},
  \bibinfo {author} {\bibfnamefont {M.}~\bibnamefont {Paternostro}}, \bibinfo
  {author} {\bibfnamefont {A.~A.}\ \bibnamefont {Geraci}}, \bibinfo {author}
  {\bibfnamefont {P.~F.}\ \bibnamefont {Barker}}, \bibinfo {author}
  {\bibfnamefont {M.~S.}\ \bibnamefont {Kim}},\ and\ \bibinfo {author}
  {\bibfnamefont {G.}~\bibnamefont {Milburn}},\ }\bibfield  {title} {\bibinfo
  {title} {Spin entanglement witness for quantum gravity},\ }\href
  {https://doi.org/10.1103/PhysRevLett.119.240401} {\bibfield  {journal}
  {\bibinfo  {journal} {Phys. Rev. Lett.}\ }\textbf {\bibinfo {volume} {119}},\
  \bibinfo {pages} {240401} (\bibinfo {year} {2017})}\BibitemShut {NoStop}%
\bibitem [{\citenamefont {van~de Kamp}\ \emph {et~al.}(2020)\citenamefont
  {van~de Kamp}, \citenamefont {Marshman}, \citenamefont {Bose},\ and\
  \citenamefont {Mazumdar}}]{Anupam02}%
  \BibitemOpen
  \bibfield  {author} {\bibinfo {author} {\bibfnamefont {T.~W.}\ \bibnamefont
  {van~de Kamp}}, \bibinfo {author} {\bibfnamefont {R.~J.}\ \bibnamefont
  {Marshman}}, \bibinfo {author} {\bibfnamefont {S.}~\bibnamefont {Bose}},\
  and\ \bibinfo {author} {\bibfnamefont {A.}~\bibnamefont {Mazumdar}},\
  }\bibfield  {title} {\bibinfo {title} {Quantum gravity witness via
  entanglement of masses: Casimir screening},\ }\href
  {https://doi.org/10.1103/PhysRevA.102.062807} {\bibfield  {journal} {\bibinfo
   {journal} {Phys. Rev. A}\ }\textbf {\bibinfo {volume} {102}},\ \bibinfo
  {pages} {062807} (\bibinfo {year} {2020})}\BibitemShut {NoStop}%
\bibitem [{\citenamefont {Marletto}\ and\ \citenamefont
  {Vedral}(2017)}]{Velatko01}%
  \BibitemOpen
  \bibfield  {author} {\bibinfo {author} {\bibfnamefont {C.}~\bibnamefont
  {Marletto}}\ and\ \bibinfo {author} {\bibfnamefont {V.}~\bibnamefont
  {Vedral}},\ }\bibfield  {title} {\bibinfo {title} {Gravitationally induced
  entanglement between two massive particles is sufficient evidence of quantum
  effects in gravity},\ }\href {https://doi.org/10.1103/PhysRevLett.119.240402}
  {\bibfield  {journal} {\bibinfo  {journal} {Phys. Rev. Lett.}\ }\textbf
  {\bibinfo {volume} {119}},\ \bibinfo {pages} {240402} (\bibinfo {year}
  {2017})}\BibitemShut {NoStop}%
\bibitem [{\citenamefont {Sinha}\ and\ \citenamefont
  {Subasi}(2020)}]{CP-coh001}%
  \BibitemOpen
  \bibfield  {author} {\bibinfo {author} {\bibfnamefont {K.}~\bibnamefont
  {Sinha}}\ and\ \bibinfo {author} {\bibfnamefont {Y.}~\bibnamefont {Subasi}},\
  }\bibfield  {title} {\bibinfo {title} {Quantum brownian motion of a particle
  from casimir-polder interactions},\ }\href
  {https://doi.org/10.1103/PhysRevA.101.032507} {\bibfield  {journal} {\bibinfo
   {journal} {Phys. Rev. A}\ }\textbf {\bibinfo {volume} {101}},\ \bibinfo
  {pages} {032507} (\bibinfo {year} {2020})}\BibitemShut {NoStop}%
\bibitem [{\citenamefont {Martinetz}\ \emph {et~al.}(2022)\citenamefont
  {Martinetz}, \citenamefont {Hornberger},\ and\ \citenamefont
  {Stickler}}]{CP-coh002}%
  \BibitemOpen
  \bibfield  {author} {\bibinfo {author} {\bibfnamefont {L.}~\bibnamefont
  {Martinetz}}, \bibinfo {author} {\bibfnamefont {K.}~\bibnamefont
  {Hornberger}},\ and\ \bibinfo {author} {\bibfnamefont {B.~A.}\ \bibnamefont
  {Stickler}},\ }\bibfield  {title} {\bibinfo {title} {Surface-induced
  decoherence and heating of charged particles},\ }\href
  {https://doi.org/10.1103/PRXQuantum.3.030327} {\bibfield  {journal} {\bibinfo
   {journal} {PRX Quantum}\ }\textbf {\bibinfo {volume} {3}},\ \bibinfo {pages}
  {030327} (\bibinfo {year} {2022})}\BibitemShut {NoStop}%
\bibitem [{\citenamefont {D'Angelis}\ \emph {et~al.}(2019)\citenamefont
  {D'Angelis}, \citenamefont {Pinheiro},\ and\ \citenamefont
  {Impens}}]{CP-coh003}%
  \BibitemOpen
  \bibfield  {author} {\bibinfo {author} {\bibfnamefont {F.~M.}\ \bibnamefont
  {D'Angelis}}, \bibinfo {author} {\bibfnamefont {F.~A.}\ \bibnamefont
  {Pinheiro}},\ and\ \bibinfo {author} {\bibfnamefont {F.}~\bibnamefont
  {Impens}},\ }\bibfield  {title} {\bibinfo {title} {Decoherence and collective
  effects of quantum emitters near a medium at criticality},\ }\href
  {https://doi.org/10.1103/PhysRevB.99.195451} {\bibfield  {journal} {\bibinfo
  {journal} {Phys. Rev. B}\ }\textbf {\bibinfo {volume} {99}},\ \bibinfo
  {pages} {195451} (\bibinfo {year} {2019})}\BibitemShut {NoStop}%
\bibitem [{\citenamefont {Fedida}\ and\ \citenamefont
  {Serafini}(2023)}]{Con01}%
  \BibitemOpen
  \bibfield  {author} {\bibinfo {author} {\bibfnamefont {S.}~\bibnamefont
  {Fedida}}\ and\ \bibinfo {author} {\bibfnamefont {A.}~\bibnamefont
  {Serafini}},\ }\bibfield  {title} {\bibinfo {title} {Tree-level entanglement
  in quantum electrodynamics},\ }\href
  {https://doi.org/10.1103/PhysRevD.107.116007} {\bibfield  {journal} {\bibinfo
   {journal} {Phys. Rev. D}\ }\textbf {\bibinfo {volume} {107}},\ \bibinfo
  {pages} {116007} (\bibinfo {year} {2023})}\BibitemShut {NoStop}%
\bibitem [{\citenamefont {Cheng}\ \emph {et~al.}(2018)\citenamefont {Cheng},
  \citenamefont {Yu},\ and\ \citenamefont {Hu}}]{Con02}%
  \BibitemOpen
  \bibfield  {author} {\bibinfo {author} {\bibfnamefont {S.}~\bibnamefont
  {Cheng}}, \bibinfo {author} {\bibfnamefont {H.}~\bibnamefont {Yu}},\ and\
  \bibinfo {author} {\bibfnamefont {J.}~\bibnamefont {Hu}},\ }\bibfield
  {title} {\bibinfo {title} {Quantum fluctuations of spacetime generate quantum
  entanglement between gravitationally polarizable subsystems},\ }\href
  {https://doi.org/10.1140/epjc/s10052-018-6433-5} {\bibfield  {journal}
  {\bibinfo  {journal} {Eur. Phys. J. C}\ }\textbf {\bibinfo {volume} {78}},\
  \bibinfo {pages} {954} (\bibinfo {year} {2018})}\BibitemShut {NoStop}%
\end{thebibliography}%


\providecommand{\noopsort}[1]{}\providecommand{\singleletter}[1]{#1}%
\providecommand{\newblock}{}

\end{document}